\documentclass[sigconf,nonacm]{acmart}%
\usepackage{caption}%
\usepackage{subcaption}%
\setcopyright{none}%
\copyrightyear{2025}%
\acmYear{2025}%
\newcommand\todo[1]{{\color{black}{#1}}}%
\newcommand\transition[1]{{\color{black}{#1}}}%
\begin{document}

\title[Quo Vadis, HCOMP?]{Quo Vadis, HCOMP?
% A
%     Review of
%     % Retrospective on
%     Twelve Years % of Research
%     of the Past
%     to Inform
%     the Future
%     of 
%     Human Computation and Crowdsourcing
    A Review of 12 Years of Research at the Frontier of Human Computation and Crowdsourcing
}%
\author{Jonas Oppenlaender}
\email{jonas.oppenlaender@oulu.fi}
\orcid{0000-0002-2342-1540}
\affiliation{%
  \institution{University of Oulu}
  \city{Oulu}
  \country{Finland}
}

\author{Ujwal Gadiraju}
\email{U.K.Gadiraju@tudelft.nl}
\orcid{0000-0002-6189-6539}
\affiliation{%
  \institution{Delft University of Technology}
  \city{Delft}
  \country{The Netherlands}
}

\author{Simo Hosio}
\email{simo.hosio@oulu.fi}
\orcid{0000-0002-9609-0965}
\affiliation{%
  \institution{University of Oulu}
  \city{Oulu}
  \country{Finland}
}


\begin{abstract}%
The field of human computation and crowdsourcing has historically studied how tasks can be outsourced to humans.
% All research fields evolve. The Recent rapid advances in artificial intelligence and the ways it is applied are now \todo{causing} shifts in the field of human computation and crowdsourcing.
However, many tasks previously distributed to human crowds can today be completed by generative AI with human-level abilities, and concerns about crowdworkers increasingly using language models to complete tasks are surfacing. 
These developments % can
    undermine core premises of the field. % itself.
In this paper, we examine the evolution of the Conference on Human Computation and Crowdsourcing (HCOMP)---a representative example of the field as one of its key venues---through the lens of Kuhn's paradigm shifts.
We review 12 years of research at HCOMP, mapping the evolution of HCOMP's research topics and identifying significant shifts over time.
Reflecting on the findings through the lens of Kuhn's paradigm shifts, we suggest that these shifts do not constitute a % full
    paradigm shift.
Ultimately, our analysis of gradual topic shifts over time, combined with data on the evident overlap with related venues, contributes a data-driven perspective to the broader discussion about the future of % the venue
HCOMP and the field as a whole.
\end{abstract}%
\begin{teaserfigure}%
\centering%
  \includegraphics[width=\textwidth]{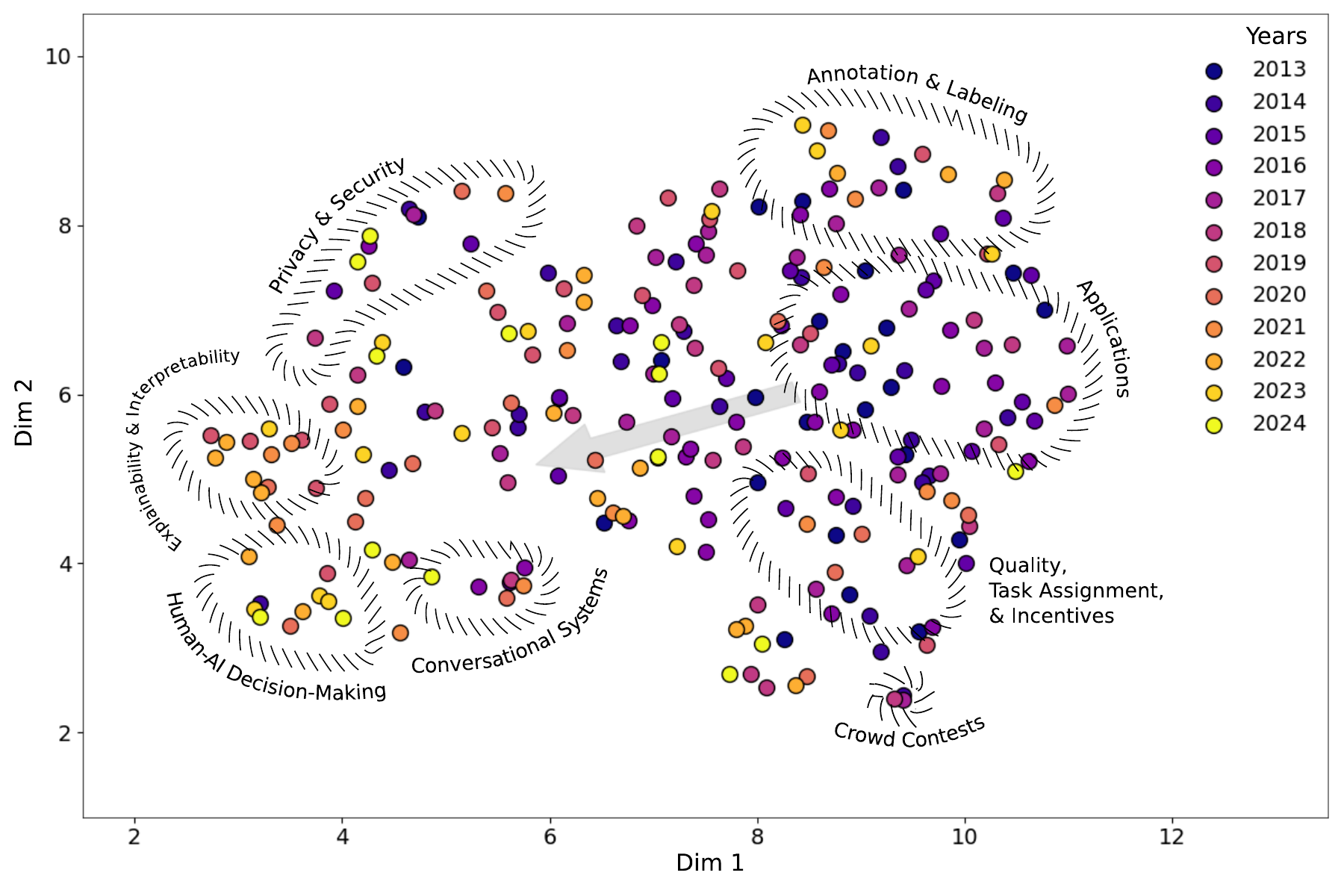}\\
  \caption{%
  Research topics in articles published at the Conference on Human Computation and Crowdsourcing (HCOMP) between 2013 and 2024.
  The dots represent article titles embedded using sentence transformers and projected into two-dimensional space with a dimensionality reduction technique (UMAP).
  The arrow indicates the general direction of the HCOMP Conference from 2013 to 2024 (centroid to centroid).
  Key themes from 2013 and 2024 are annotated, demonstrating how many articles in HCOMP have migrated away from HCOMP's traditional key motor themes (such as annotation \& labeling, quality, incentives \& task assignment, and applications) toward the topics of explainable AI (XAI), conversational systems, and human-AI decision-making.
  % To create this visualization, we encoded the titles of HCOMP articles and UMAP-projected the resulting embeddings into 2D space, with
  % Clusters denote research topics.
  An interactive visualization is available at
    %%% \textbf{[ANON-LINK]}.
    \href{https://hcomp-retrospective.github.io/}{https://hcomp-retrospective.github.io}.
  }%
  \Description{A visualization of embeddings denoting clusters of research topics.}
  \label{fig:teaser}%
\end{teaserfigure}%
\maketitle
%%%%%%%%%%%%%%%%%%%%%%%%%%%%%%%%%%%%
%%%%%%%%%%%%%%%%%%%%%%%%%%%%%%%%%%%%
%
% ======================
\section{Introduction}%

The field of human computation and crowdsourcing has long relied on harnessing human ingenuity to address complex problems.
Foundational work---such as Luis von Ahn’s early contributions with projects such as the ESP game and CAPTCHA \cite{Ahn,ESP.pdf}---established a paradigm for leveraging % on-demand
    human input.
    % for the field.
    % Crowdsourcing emerged as a
    %     % fruitful
    %     productive
    % research area, applying and studying the practice of outsourcing a task to a crowd of people \cite{COIN}.
    Crowdsourcing~\cite{COIN} emerged as a productive research field, exploring % both
    the theoretical and practical dimensions of distributing tasks to a crowd.
Over the years, the field evolved through what can be seen as a period of ``normal science''~\cite{kuhn1997structure}, focused on solving fundamental issues in crowdsourcing---optimizing task design and incentives, ensuring data quality, exploring novel workflows, and refining models of human interaction---all while operating within a well-defined set of assumptions and methods~\cite{law2011human,demartini2017introduction,kittur2013future}.

All scientific fields evolve, adapting % They adapt 
to new developments and emerging challenges. In recent years, however, the rapid progress of artificial intelligence has begun to shake % some of 
the foundations of fields concerned with human input, labor, and cognition.
Human computation and crowdsourcing is one such field.
Tasks once assigned to human workers can now be performed at least partially by large language models, raising questions about the very role of human input in crowdsourcing~\cite{wu2023llms}.
Concerns have also emerged that crowdworkers may be increasingly relying on automated tools to complete tasks, potentially undermining core premises of human computation~\cite{veselovsky2023artificial,faggioli2023perspectives}.
%Experimental findings have uncovered anomalies, and 
Other related developments, such as data-labeling firms rebranding as AI companies, further continue to disrupt the established framework of crowdsourcing.
% While HCOMP encompasses more than just crowdsourcing,
These developments offer a timely % and exciting
opportunity to revisit the field's scope, assumptions, and to envision its possible future directions.

In this paper, we investigate shifts at the Conference on Human Computation and Crowdsourcing (HCOMP) as a proxy into the wider field of research on human computation and crowdsourcing.
We adopt Kuhn’s notion of paradigm shifts~\cite{kuhn1997structure} as a lens to examine the evolution of the HCOMP conference.
    Kuhn’s model characterizes scientific progress as a series of distinct phases. In the pre-science phase, a field lacks consensus, and diverse, often conflicting theories coexist. This is followed by a period of normal science, during which a dominant paradigm emerges and research focuses on solving puzzles within that established framework. As anomalies and unexpected findings accumulate, the field may enter a crisis, challenging the core assumptions of the prevailing paradigm. If these challenges cannot be reconciled, a revolutionary phase occurs, leading to a paradigm shift in which the old framework is replaced by a new one that redefines the discipline.
% Kuhn argues that a true paradigm shift involves a fundamental reconceptualization of a field’s underlying principles rather than merely a cumulative improvement of existing methods.

% One could argue that redefining human input may itself be a revolutionary shift.
One could argue that the role of human input being redefined already constitutes a revolutionary phase.
%Following such a crisis would be a revolutionary phase---one in which, for instance, the very role of human input is being redefined.
The apparent challenges---brought about by the disruptive influence of generative AI---further suggest a form of incommensurability between the established paradigm and the new realities imposed by large language models, potentially leading to non-linear progress that defies past metrics of evaluation.
Moving from ``normal science'' to revolutionary science requires questioning the fundamental aspects of the field (such as the need for human input) and an exploration of alternatives.
And a true paradigm shift involves the fundamental reconceptualization of a field’s underlying principles rather than merely a cumulative improvement of existing methods.
It is worth questioning whether there has % already
been a paradigm shift at HCOMP, or whether we are merely witnessing a gradual, natural shift in topics.

Our work undertakes a detailed analysis of  research published at the HCOMP conference with
% , the field’s % flagship
% key venue.
% We adopt 
a multi-method approach, to capture both the historical evolution and emerging trends within the community.
We begin by employing embedding techniques and clustering algorithms to map research topics and identify shifts over time.
%%% To capture the tone of scholarly contributions, we conduct an article-level sentiment analysis.
%%% We also evaluate the influence of HCOMP articles by analyzing citation impact via Google Scholar data.
Further, we compare the HCOMP conference with six related conferences by measuring the cosine similarity of article title embeddings, which allows us to speculate on the future trajectory of the field.
This is complemented by a co-word analysis that examines the relationships among key terms at HCOMP and across conferences. % proceedings.
Finally, we measure shifts in research topics at HCOMP with the aim of identifying whether a paradigm shift has taken place at HCOMP.
    % and what phase the field is in,
    % according to Kuhn's model of paradigm shifts.
Together, our % multi-method
analysis provides a comprehensive view of the evolution of % fundamental shifts occurring within the field of
HCOMP, illuminating both the enduring strengths of the traditional paradigm during the period of ``normal science'' and the recent disruptive challenges introduced by generative AI.

We contribute:%
\begin{itemize}%
    % \item A retrospective and critical review of crowdsourcing's early history.
    %     %, highlighting flaws in the paid microtask crowdsourcing paradigm.
    \item An empirical investigation into the evolution of the HCOMP conference, the key venue for research on human computation and crowdsourcing.
    We highlight recent developments and fundamental shifts in the conference's research topics and analyze co-occurring words. % and author sentiment.
    \item An investigation of shifts at the HCOMP conference in relation to six related conferences---Collective Intelligence (CI),  CSCW,  FAccT,  IUI,  UMAP, and AAMAS---providing valuable information to inform the future of HCOMP.
    %\item We contribute an online visualization for interactively exploring the conference's evolution in research topics. %, and for exploring the HCOMP's relation to six other related conferences.
    \item A discussion of these findings through the lens of Kuhn's model of paradigm shifts.
    Our work can help inform others wishing to analyze the evolution of research at scientific venues in a similar way.
\end{itemize}%

By framing the discussion in terms of a potential paradigm shift, we explore the critical juncture in the evolution of human computation and crowdsourcing,
marked by the transformative impact of generative AI.
This perspective highlights the crisis of reconciling traditional methods with new technological capabilities and invites a broader discussion on the future direction of the field.

\section{Related Work}%
\label{sec:related-work}%
% ======================

% ----------------------
\subsection{Kuhn’s Paradigm Shifts}
\label{sec:kuhn}
% ----------------------
Kuhn’s model of scientific progress
    %, as outlined in ``The Structure of Scientific Revolutions''
\cite{kuhn1997structure} offers a framework for understanding how disciplines evolve through four distinct phases:%
\begin{enumerate}%
    \item 
\textit{Pre-science}:
In this initial phase, a field lacks a unified theoretical framework. Researchers pursue diverse and often conflicting approaches without a shared set of standards or observational criteria. This period is characterized by debates over fundamentals, where as many theories exist as there are theorists.
    \item 
\textit{Normal science}:
Once a dominant paradigm is established, the field enters a phase of normal science. Researchers work within this established framework, addressing puzzles and refining existing methods rather than challenging the core assumptions. Anomalies—observations that do not easily fit the paradigm—are typically treated as challenges to be solved within the current structure, rather than reasons to question it.
    \item 
\textit{Crisis}:
Over time, if anomalies accumulate and prove resistant to resolution, confidence in the established paradigm begins to wane. This phase is marked by a growing sense of crisis, as the foundational assumptions of the field are increasingly questioned. Researchers start to explore alternative explanations, and competing theories emerge to address the persistent anomalies.
    \item 
\textit{Revolution}:
Should the crisis remain unresolved, the field may undergo a revolutionary shift. In this phase, a new paradigm emerges—one that redefines the field’s basic principles and methods. The new framework is not simply an extension of the old one but represents a fundamental change in how problems are understood and approached. Kuhn emphasizes that this shift is driven by both empirical findings and sociological factors, making the transition complex and non-linear.
\end{enumerate}

%%% Examples:
Kuhn's model % stems from the context natural science, it is also applicable to
has been applied in % the context of
computer science. % before.
    For instance, the model was used to frame developments in the field of computer vision, where researchers % were 
    eagerly % to
    adopted % the significant
    advances in deep learning \cite{kriegler2022paradigmatic}.
    %%%
    In other fields and scientific disciplines, deep learning has also had a strong impact, enabling new ways of science \cite{bianchini2020deeplearningscience}.
    %%%
    Another example is prompt-based learning (i.e., prompting large pre-trained models), which brought paradigm shifts in the fields of AI and Natural Language Processing \cite{10.1145/3560815}.
    %%%
    In Human-Computer Interaction, using synthetic participants (e.g., for usability testing) is a growing trend \cite{3708815.pdf}. Simulating users with generative AI is a new frontier that fundamentally challenges the traditional assumption that HCI studies must involve human participants \cite{3708815.pdf,schmidt2024simulating,3613904.3642703.pdf}.
    %%%
    Another example is, arguably, education which is undergoing a shift brought about by generative AI \cite{10.3389/feduc.2023.1161777}.

Kuhn’s model provides a useful lens through which to view the evolution of research in crowdsourcing and human computation.
\transition{In the following section, we provide a % brief
    retrospective on HCOMP's phase of ``normal science.''}

% ----------------------
\subsection{{A Retrospective on} HCOMP's Period of ``Normal Science''}
\label{sec:topics}%
% ----------------------
During HCOMP's phase of normal science, several research topics served as key motor themes for the field. Early work focused on quality control as a fundamentally important aspect of human computation and crowdsourcing. The quality of crowdsourced responses was found to be a critical bottleneck in many applications. %Quality was a topic of study from the very beginning of the research field.
% The humans working in online crowdsourcing platforms are % faceless---
%     an anonymous crowd, completing tasks as if they were a miraculous human-powered automaton.
    As early as in the year \citeyear{1357054.1357127.pdf}, \citeauthor{1357054.1357127.pdf}  noted in their crowdsourced user studies that almost 50\% of the responses on Amazon Mechanical Turk ``consisted of uninformative responses including semantically empty [...], non-constructive [...], or copy-and-paste responses'' \cite{1357054.1357127.pdf}.
        %% "Out of the total of 210 free-text responses regarding how the article could be improved, 102 (48.6%) consisted of uninformative responses including semantically empty (e.g., “None”), non-constructive (e.g., “well written”), or copy-and-paste responses (e.g., “More pictures to break up the text” given for all articles rated by a user)."
%
This wasteful ratio of good to bad responses persisted over the years, with quality-control methods being proposed to overcome existing challenges. 
This paved way for the use of 
    gold-standard questions~\cite{oleson2011programmatic},
    post-hoc filtering~\cite{DanielCSUR2017.pdf},
    statistical and algorithmic methods to control for quality~\cite{baba2013statistical},  collusion detection \cite{KhudaBukhsh_Carbonell_Jansen_2014}, 
    pre-task worker selection  
    and behavior-based quality control methods~\cite{feldman2014behavior}
    emerging as key methods for improving response quality.
    %after data collection.
%
Approaches from psychology and survey research, such as Instructional Manipulation Checks (IMC) \cite{OPPENHEIMER2009867}, were adopted by the field of crowdsourcing. Over time, crowd workers adapted to the evolving quality control measures % being used for quality control
and were found to be more attentive to IMC than other human subject pools \cite{Hauser2016},
and
\citeauthor{Checco_Bates_Demartini_2018} later demonstrated how 
gold questions % is prone to attacks
can be gamed \cite{Checco_Bates_Demartini_2018}.
There was also a growing interest in quality control within citizen science initiatives, exploring a different set of intrinsic incentives for % worker
participation~\cite{jennett2014eight,wiggins2014sensor,bowyer2015panoptes}.

% \todo{[TODO(UG): Caring about worker experiences (beginning form 2014) \cite{bigham2015s}]}

Leveraging crowdsourcing methods to address real-world problems and use cases was another strong research stream at HCOMP during the period of ``normal science.'' Applications included, for instance, the synthesis of information \cite{Luther_Hahn_Dow_Kittur_2015},
    paper screening for literature reviews \cite{Krivosheev_Casati_Caforio_Benatallah_2017},
    augmenting video \cite{Salisbury_Stein_Ramchurn_2015},
    conference scheduling \cite{Bhardwaj_Kim_Dow_Karger_Madden_Miller_Zhang_2014},
    and a genomics game \cite{Singh_Ahsan_Blanchette_Waldispühl_2017}.
Annotation and labeling was another large area of focus at HCOMP during this period (see \autoref{fig:teaser}), % top right),
with research on methods and algorithms for aggregating labels to fuel training of computer vision models.
Notably, work by~\citeauthor{sheshadri2013square} to improve response aggregation methods in crowdsourcing was impactful~\cite{sheshadri2013square}.
The authors presented an open source shared task framework including benchmark datasets,
defined tasks, standard metrics, and reference implementations with empirical results for
 popular methods at the time.
 % Other works addressed aggregation of boolean tasks \cite{Alfaro_Polychronopoulos_Shavlovsky_2015}.

While there has been a strong focus on microtasks at HCOMP, applications in alternative areas were also explored, such as citizen science \cite{Xue_Dilkina_Damoulas_Fink_Gomes_Kelling_2013}, crowdfunding \cite{Horvát_Wachs_Wang_Hannák_2018}, and crowd contests \cite{Cavallo_Jain_2013,Sarne_Lepioshkin_2017}.
%A sub-category among applications is 
We also observed
    % creative
    % location-based
    geo-enabled 
applications such as spatial crowdsourcing and crowdsensing, with notable examples such as earthquake detection using citizen science \cite{Flores_Guzman_Poblete_2017},
    local crowds for event reporting \cite{Agapie_Teevan_Monroy-Hernández_2015},
    and participatory sensing \cite{Zenonos_Stein_Jennings_2017}.
Real-time applications started becoming a topical focus at HCOMP in 2016, including works on
    real-time question-answering \cite{Savenkov_Agichtein_2016},
    real-time disease information \cite{Mutembesa_Omongo_Mwebaze_2018}, and
    real-time assistance in real environments \cite{EURECA,abbas2021making}.
    % There was also a growing interest in tackling challenges in  conversational systems  ~\cite{ikeda2018utilizing,abbas2021making}.

%\todo{(UG:to finalize this subsection)}

% Crowd work quality~\cite{Jung_Lease_2015} and
Workflow and task design have also received strong attention in the HCOMP community~\cite{Wu_Quinn_2017,Manam_Quinn_2018,Goto_Ishida_Lin_2016}.
Cost-quality-time optimization~\cite{Goel_Rajpal_Mausam_2017}, predicting label quality~\cite{Jung_Park_Lease_2014}, or aggregation mechanisms~\cite{Waggoner_Chen_2014} were some objectives pursued in this direction.
%
% From a more worker-centric perspective,
Task routing and incentive design have received keen interest, too.
For instance, parallelization of tasks~\cite{Bragg_Kolobov_Mausam_Weld_2014}, skill and stress aware task assignment~\cite{Kumai_Matsubara_Shiraishi_Wakatsuki_Zhang_Shionome_Kitagawa_Morishima_2018}, and dynamic task assignment to crowd workers versus AI~\cite{kobayashi2021human} have been explored.
%Worker incentives matter too; 
Different pricing schemes~\cite{Difallah_Catasta_Demartini_Cudré-Mauroux_2014} or incentives to increase engagement and counter bias~\cite{Faltings_Jurca_Pu_Tran_2014} have been explored. Monetary interventions were utilized to prevent task switching~\cite{Yin_Chen_Sun_2014} and to predict work quality~\cite{Yin_Chen_2016}.

In the years that followed, 
    % there was work that 
    % Later research 
    researchers
explored agreement and disagreement mechanisms
\cite{Checco_Roitero_Maddalena_Mizzaro_Demartini_2017}, linguistic frame disambiguation \cite{Dumitrache_Aroyo_Welty_2018}, and the use of dummy events to improve worker engagement \cite{Elmalech_Sarne_David_Hajaj_2016}. Others explored training workers and leveraging worker skills in different contexts, such as providing stress management support~\cite{abbas2020trainbot}, or music annotation~\cite{samiotis2021exploring}, and developed methods to ensure fair wages~\cite{whiting2019fair} or support novice workers~\cite{rechkemmer2020motivating}. 
Over the years, efforts have also been invested to understand crowd worker behavior---including workers' strategies to maximize earnings~\cite{kaplan2018striving}, their goal-setting behavior~\cite{abbas2022goal}---and improve worker experiences in different contexts~\cite{das2020fast,hettiachchi2020context} and worker communities \cite{Zaamout_Barker_2018,Valk_Hoßfeld_Redi_Bozzon_2018}. Others explored alternative input modalities to lower the barrier for participation in crowd work~\cite{singh2022signupcrowd,allen2022gesticulate}.

\transition{%
In this paper, we investigate how research in the field of human computation and crowdsourcing has shifted from ``normal science'' to a new phase over the past twelve years.
% The following section describes our multi-method approach.%
To do so, we adopt a multi-method approach, which we detail in the following section.%
}%
%
%
%
% ======================
\section{Method}%
\label{sec:method}%
% ======================
%
\transition{%
We analyze % topical
shifts at the HCOMP conference
    % and related conferences
from multiple perspectives through the lens of Kuhn's model.
% , based on review of the literature and a linguistic analysis of articles published at the Conference on Human Computation and Crowdsourcing (HCOMP).
In the following, we describe our data collection and analysis.%
}%
%
%
% ----------------------
\subsection{Data Collection}%
% ----------------------
% The proceedings of the HCOMP conference comprise a total of 250 % full
% research articles between 2013 and 2024.
% The proceedings contain between 14 and 27 articles ($\textrm{Mean}=20.8$, $\textrm{SD}=4.3$).
% We scraped the titles and abstracts of all articles % ($N=250$)
% % published at the HCOMP conference  
% from the Association for the Advancement of Artificial Intelligence (AAAI)'s website.
We collected the titles and abstracts of all research articles ($N=250$) published at the HCOMP conference % proceedings
from 2013 to 2024. %, comprising the complete set of 250 papers.
Each year’s proceedings include between 14 and 27 articles ($\textrm{Mean}=20.8$, $\textrm{SD}=4.3$).
The data was scraped from the website of the Association for the Advancement of Artificial Intelligence (AAAI).
% For each article, we combined the title and abstract into a single document.
% We then removed stopwords from the documents.
%     Stopwords are words in the English language that are not informative since they are very common across the corpus (e.g., the, of, and, etc.).
% \todo{We applied stemming to all remaining terms, while keeping a full list of the original terms belonging to each stem.}
% The documents contain between 98 and 485 tokens ($M=210.7$ tokens) and are our unit of analysis, if not stated otherwise.
% The documents contain between 98 and 485 tokens ($M=210.7$ tokens).
% Min length (tokens): 98
% Max length (tokens): 485
% Mean length (tokens): 210.672
%
%
%%% <descriptive stats aBOUT TITLES to fill space>
    % Mean articles per year: 20.83
    % Standard deviation (articles per year): 4.34
    % Minimum articles per year: 14
    % Maximum articles per year: 27
    % Median articles per year: 21.5
% The % mean number of characters in the article titles is 77.1 ($\textrm{Min}=26$, $\textrm{Max}=138$, $\textrm{Median}=74$),
    % Min title length: 26
    % Max title length: 138
    % Mean title length: 77.148
    % Median title length: 74.0
% or an 
% titles contain, on average, 10.7 tokens ($\textrm{Min}=3$, $\textrm{Max}=28$, $\textrm{Median}=10$).
    % Min title tokens length: 3
    % Max title tokens length: 28
    % Mean title tokens length: 10.66
    % Median title tokens length: 10.0
Title lengths range from 3 to 28 tokens ($\textrm{Mean}=10.7$, $\textrm{Median}=10$). % , with an average of 10.7 and a median of 10.
\transition{%
The collected data was analyzed using multiple methods, as described below.
}%
%
%
% ----------------------
\subsection{Data Analysis}%
\subsubsection{Initial exploration}%
\label{sec:method:proceedings-review}%
% ---
The lead author started to explore the proceedings of the HCOMP conference to develop an overall understanding of the venue by using Voyant Tools \cite{voyant}.
% This was complemented by an exploration with Voyant-Tools \cite{voyant} to explore common research topics across the entire corpus.
We then proceeded to review works published at HCOMP, focusing on titles and abstracts, to identify research themes and topics at the conference.
% The HCOMP proceedings comprise 250~research articles (2013--2024).
This exploration and review informed sections \ref{sec:topics} and \ref{sec:topics:shift}.%
%
%
% ---
\subsubsection{Topic analysis}%
\label{sec:method:topic-analysis}%
To identify relationships between topics, we encoded the article titles % (including stopwords)
into embeddings using Sentence Transformers \cite{reimers-gurevych-2019-sentence} (all-mpnet-base-v2) 
and used UMAP \cite{UMAP} to project the embeddings into a two-dimensional space.
% reduced the embedding dimensions into two-dimensional space with UMAP \cite{UMAP}, 
UMAP is a dimensionality reduction technique which preserves local and global structures better than t-SNE and PCA~\cite{Understanding-UMAP,UMAP}.
    % The sentence transformer captures contextual relationships, word order, and deeper semantics. %, making it more robust than simple keyword matching.
    % Therefore, the embedding space reflects the semantic similarity of article titles, and titles with a similar meaning will appear close to each other in embedding space.
    The sentence transformer captures contextual relationships, word order, and deeper semantics.
    As a result, the embedding space reflects semantic similarity: titles with similar meanings are positioned closer together.
% The embedding space reflects the similarity of paper titles, based on semantic meaning.
We used clustering to identify the approximate locations of topics in embedding space
% To this end, we 
by iteratively applying HDBSCAN~\cite{HDBSCAN}, a density-based clustering method, with different parameters.
% and determine the optimal minimum cluster size empirically as 5.
% Note that this is not 100\% accurate, due to different clustering parameters leading to different results.
The exploration of different clustering solutions allowed us to get an overview of the structure of the embedding space and the trends within.
% For instance, some documents remained unclustered due to these parameter settings, such as the small cluster related to crowdfunding.
We manually annotated the clusters and indicate the general trend with an arrow, which we calculated from the embedding centroids of the 2013 and 2024 HCOMP proceedings.
    % The centroid is calculated by taking the mean of all embeddings in a given year.
    The centroid is the `mean embedding' (i.e., the point in space that, on average, is closest to all other data points in a given year).
% track publication years to 
Further, we mapped how HCOMP topics, as identified by the clustering algorithm, have evolved over time (see figures \ref{fig:teaser}, \ref{fig:conferences}, and \ref{fig:conf}).%
\subsubsection{Paradigm shift}%
% ---
We use the notion of a Gestalt-shift in the context of Kuhn's framework to measure whether a sudden shift in research topics has taken place at the HCOMP conference.
%
% The idea is that by measuring the % year-by-year
% cosine distance between the embedding centroids of article titles at the HCOMP conference in subsequent years, we can measure whether a shift in research topics has taken place, and we can quantify when it has taken place.
% If there was a strong increase in the cosine distance between centroids in subsequent years, we could assert that a sudden shift has taken place.
The idea is to measure the cosine distance between the embedding centroids of article titles % from the HCOMP conference
across consecutive years.
This allows us to assess whether a shift in research topics has occurred, and when it took place.
A sharp increase in cosine distance between centroids from one year to the next would suggest a sudden shift in research focus.
Of course, whether a detected shift constitutes a paradigm shift is arguable, since it is not clear what magnitude of shift would constitute a paradigm shift.
    Or in other words, how far would the HCOMP conference need to move away from its traditional research topics to constitute a paradigm shift?
%We cannot answer this question in this paper.
Given how the HCOMP conference is affected by recent developments in AI,  %Generative AI can now solve many of the classical crowdsourcing tasks, including image segmentation and classification.
we expect there to be a % measurable
notable shift in research topics in recent years.
The results are depicted in figures \ref{fig:teaser}, \ref{fig:cosine-distances}, \ref{fig:conf}, and~\ref{fig:gestalt}.%
%
%
%
% % ---
% \subsubsection{Sentiment analysis}%
% % ---
% We % use VADER \cite{VADER} % or LIWC
% analyzed 
% sentiment in articles published at HCOMP as a proxy for %  have become more positive or negative over time.
% % This may indicate
% how authors feel about their research area.
% To investigate this, we applied VADER sentiment analysis~\cite{VADER} to the documents. % (consisting of titles and abstracts). % representing articles published at HCOMP between 2013 and 2024.
%     VADER (Valence Aware Dictionary and sEntiment Reasoner) is a lexicon- and rule-based sentiment analysis tool designed to measure the sentiment polarity of text. %, particularly in social media and informal language.
% VADER assigns scores based on the valence of words and their contextual modifiers, producing four key metrics: positive, neutral, and negative sentiment scores
% %, which sum to 1,
% and a compound score, % normalized
%     ranging from -1 (most extreme negative) to +1 (most extreme positive).
% In our analysis, we examined the compound sentiment scores over all documents per year (see \autoref{fig:sentiment}).%
%
%
%
% ---
\subsubsection{Conference analysis}%
% ---
To inform decision-making on the future of the HCOMP conference and trigger reflection in the HCOMP community, %(e.g., a potential merge of HCOMP with another conference).
we used the same approach as in Section~\ref{sec:method:topic-analysis} and encoded the titles of articles published at six related conferences from 2013--2024:
    ACM Collective Intelligence Conference (CI; $N=220$),
    ACM SIGCHI Conference on Computer-Supported Cooperative Work \& Social Computing (CSCW; $N=3,081$),
    ACM Conference on Fairness, Accountability, and Transparency (FAccT; $N=657$),
    ACM Conference on Intelligent User Interfaces (IUI; $N=612$),
    ACM Conference on User Modeling, Adaptation and Personalization (UMAP; $N=232$),
    and 
    the Conference on Autonomous Agents and Multiagent Systems (AAMAS; $N=2,203$).
% we can calculate the cosine distance (or, inversely, the cosine similarity) between the titles of articles published in these conferences.
    These conferences were selected because they have some overlap with HCOMP in the past or present. % (or, potentially, future).
Note that for ACM CI, some older proceedings were no longer accessible.
We plot the resulting embeddings into twodimensional space using UMAP.
Since embeddings are numeric vector representations of semantic meaning encoded in text,
the plots give us a topical overview of HCOMP's relation to other related conferences and the direction of the recent shift in HCOMP,
    % In other terms, we can quantitatively answer the question in which direction HCOMP moved in recent years,
    in terms of centroid cosine distance of conference proceedings.
% This allows us to quantify how the six conferences have evolved since 2013 in terms of topics studied and their similarity to the HCOMP Conference.
The results are depicted in \autoref{tab:nearest_conf}, \autoref{fig:conferences}, and \autoref{fig:conf}.
% we further analyzed shared keywords between the conferences by tokenizing the article titles and counting co-occurrences (see \autoref{fig:shared-keywords} and \autoref{fig:sankey}).

% ---
\subsubsection{Co-word analysis}%
% ---
To complement our analysis of topics and conferences, we analyzed co-words in the titles and abstracts of HCOMP articles (excluding stopwords).
    Co-words are co-occurring words that are frequently used together in a sentence.
% We counted the frequency of occurrence of co-words insensitive to the order of terms in the co-word pair.
% We plotted the frequency of co-words at the HCOMP conference over time, including only those co-words that appear in more than one year (see \autoref{fig:cowords}).
% \autoref{fig:shared-keywords},
We counted the frequency of co-word pairs, treating them as unordered (i.e., ignoring the order of terms in a co-word pair).
We then plotted the frequency of these co-words at the HCOMP conference over time, including only those that appeared in more than one year (see \autoref{fig:cowords}).
Further, we compared shared keywords in the titles of articles at  HCOMP  and the six related conferences (see figures \ref{fig:shared-keywords} and \ref{fig:sankey}).%

% ======================
\section{Results}%
\label{sec:results}%
\subsection{Recent Shift in Topics}
\label{sec:topics:shift}%
% ---
The initial years of HCOMP, as discussed earlier, were focused on optimizing and addressing issues around crowd work,
    % (such as quality, task design, and work flows), 
    but also applications of crowdsourcing.
Since 2018, we can identify a gradual
    % transition at HCOMP to a new phase, as evident in a gradual
shift of research topics
    % and keywords used in HCOMP articles
    studied at HCOMP.
% (see figures \ref{fig:teaser}, \ref{fig:cowords}, and \ref{fig:conferences}). 

Since 2018, a gradual shift in HCOMP's focus toward tackling problems at the intersection of humans and AI systems can be clearly observed (as represented in the bottom-left of Figure~\ref{fig:teaser}%
% and \autoref{fig:topics-over-time}
). 
With the growing advances in machine learning and recognizing important societal implications, the HCOMP community began to address challenges around
    bias and fairness~\cite{Otterbacher_2018,Nushi_Kamar_Horvitz_2018,otterbacher2019we,draws2021checklist,biswas2024hi},
    interpretability~\cite{Lage_Chen_He_Narayanan_Kim_Gershman_Doshi-Velez_2019,melis2021human}, explainability~\cite{nourani2019effects, ray2019can,he2022like,liao2022connecting},
    privacy, trust and reliance on  AI systems~\cite{alshaibani2020privacy,erlei2020impact,bansal2019beyond},
    human-AI decision making~\cite{grgic2022taking,yacoby2022if,rastogi2023taxonomy,lu2024mix,Nushi_Kamar_Horvitz_2018},
    human-AI team performance \cite{bansal2019beyond},
    collaborative human-AI methods~\cite{zong2023crowd,lim2023backtrace},
    and
    AI risks~\cite{bogucka2024atlas}.

% ---
%
\begin{figure*}[!htb]%
\centering%
  \includegraphics[width=.8\textwidth]{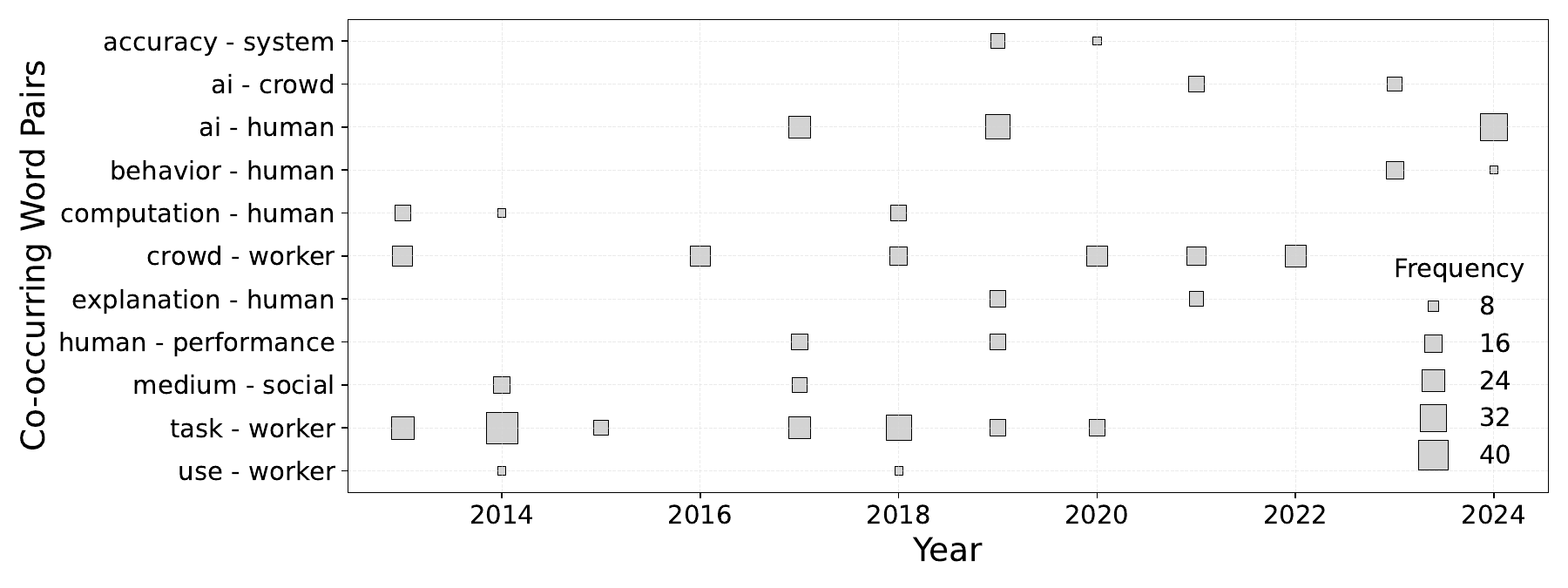}
  \caption{Co-word occurrences at the HCOMP Conference over time  (order-insensitive, % multi-year
  considering only co-words that appear in more than one year)}
  \Description{A plot showing how co-word pairs have evolved over time. It is noticeable that the pairs task-worker and crowd-worker are no longer being used in recent HCOMP proceedings. Instead, the community now focuses on AI-crowd and Human-AI.}%
  \label{fig:cowords}%
\end{figure*}%
%
% ---
This shift in research focus is also evident in our co-word analysis (see \autoref{fig:cowords}), where
the co-word pairs \textit{task--worker} and \textit{crowd--worker} ceased to be present in the HCOMP titles and abstracts in 2021 and 2023, respectively.
Instead, the HCOMP community moved to using more human-centered co-word pairs, such as \textit{human--behavior}, \textit{AI--crowd}, and \textit{human--AI}.
%     speaking to a degree of (long overdue) humanization of the crowd the field of HCOMP.
Our analysis also shows that this broadening in perspectives % toward topics at the intersection of humans and AI systems
does not coincide with the introduction of OpenAI's popular ChatGPT language model in 2022.
Instead, a reorientation is notable as early as 2019, with clear changes in co-word pairs becoming % apparent
notable one year before OpenAI introduced ChatGPT (cf. \autoref{fig:cowords}).
In that year, OpenAI released GPT-3~\cite{NEURIPS2020_1457c0d6}, a language model that with 175~billion parameters had over 100 times the size of its predecessor GPT-2.
Perhaps it was this new model that raised both interest in AI but also heightened concerns in the HCOMP community.

In 2018, the HCOMP community started to demonstrate concerns raised by increasingly intelligent automation tools.
One notable incident occurred in mid-2018, when researchers outside the HCOMP community reported
% ``bot-like'' responses and a drop in quality in crowdsourced data,
a decline in the quality of crowdsourced data, along with responses that appeared to be generated by ``bots,''
speculating that fraudulent activity and potentially automation was at play 
\cite{Evidence.pdf,Ryan.pdf,10.1177@1948550619875149.pdf,NewScientist,ABotPanicHitsAmazonMechanicalTurk_WIRED.pdf}.
    % Wired reported a ``bot panic'' among scientists using Amazon Mechanical Turk for academic studies \mbox{\cite{}ABotPanicHitsAmazonMechanicalTurk_WIRED.pdf}.
% ~\cite{SavingMTurk_Feb7.pdf,dennis2018.pdf}.
% VPS and VPS could potentially also be in some cases connected to bots.
% More recently, first evidence of scripting and automation have alerted scientists and practitioners.
In their blog posts, \citeauthor{After_the_Bot_Scare} and \citeauthor{Moving_Beyond_Bots} later concluded that % the mid-2018 issue 
this incident
% Turkprime, a recruitment platform integrated with MTurk, and found that the reported low  data quality 
was likely due to ``farmers''---i.e., workers using `server farms' for submitting HITs \cite{After_the_Bot_Scare,Moving_Beyond_Bots}.
% Perhaps a first concrete sign of such tools became evident in
In the same year, \citeauthor{Kaplan_Saito_Hara_Bigham_2018} studied work strategies and tool use among crowd workers \cite{Kaplan_Saito_Hara_Bigham_2018}.
Automation tools have, of course, been used by workers for long already, but more prominently for task management than data generation.
In the hands of crowd workers, the use of automated tools for generating answers to tasks is a threat to the % quality and
validity of data collected on crowdsourcing platforms.
    %---meant to be a product of human intelligence. Note that others have identified advances in generative AI as an opportunity to make crowdsourcing workflows more effective~\cite{allen2023power,wu2022ai}.
% While automated tools were allowed by MTurk, task automation became a concern.
%
% ``Crowd farm'' may also refer to institutions in a low-income countries which employ workers to complete tasks on crowdsourcing platforms \cite{crowdfarmschina}.
These developments highlighted growing tensions between human labor and automation on crowdsourcing platforms,
% While the causes and implications of these incidents were still being debated, they 
coinciding with a broader shift in the HCOMP community’s focus.%

% ---
% \input{FIG-TOPIC-BARS}
% ---

% ---
%
\begin{figure*}[!thb]%
\centering%
  \includegraphics[width=.8\textwidth]{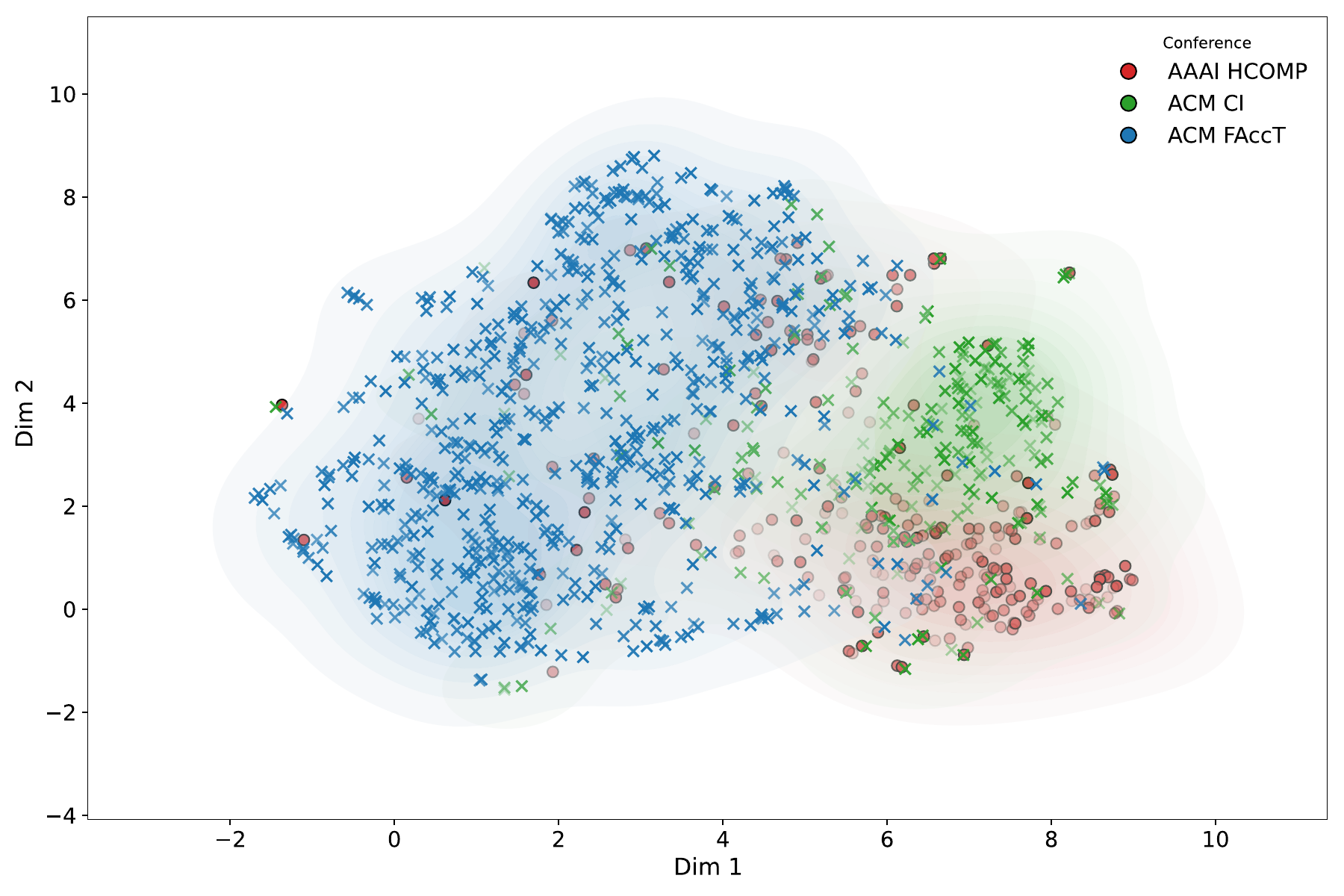}
  \caption{%
  Comparison of article titles in HCOMP, ACM Collective Intelligence (CI) and ACM Conference on Fairness, Accountability, and Transparency (FAccT).
  Titles of more recent articles are more opaque.
  }%
  \Description{A topical comparison between the HCOMP, CI, and FAccT conferences shows a migration of HCOMP authors in recent years toward FAccT.}
  \label{fig:conferences}
\end{figure*}%
%
% ---

%Nevertheless, this external incident may have % shaken up the HCOMP community, leading to
% more research on critical issues involving workers.
%lead the HCOMP community to start questioning %some of its fundamental assumptions.

% Ultimately, the field's longstanding focus on microtasks and crowd workers was almost completely dropped in the year 2021 (see \autoref{fig:cowords}).

% This could be referred to as the (long overdue, some could say) humanization of HCOMP.
% With this reorientation from workers and task efficiency to other topics came a change in vocabulary at the conference (see \autoref{fig:vocabulary}). Since 2021, the 25 most important terms, by TF-IDF, no longer included `worker'. Instead, the term `human' rose to prominence (see \autoref{fig:2013-2014}).

%%% Human-AI decision-making
% human-machine decision making and human-machine teaming have become important topics of study in recent iterations of the HCOMP Conference.
% \todo{<examples of human-AI topics studied>}

%%% conversational systems and conversational agents

%%% Explainability and interpretability
Since then, the commoditization of AI % , while introducing some of the mentioned concerns, 
has drawn interest from some members of the HCOMP community to % recently
%shift focus to 
research topics that fall within the % core
focus of other venues.
    % (see \autoref{fig:conferences}, \autoref{fig:shared-keywords}, \autoref{fig:sankey}, \autoref{fig:conf}, \autoref{fig:hulls}, and \autoref{tab:nearest_conf}).
Specifically, some recent research at HCOMP now strongly relates to topics studied at  ACM FAccT (see \autoref{fig:conferences}), a conference focusing on issues such as algorithmic transparency, fairness in machine learning, explainability and interpretability, bias, and ethics.
By cosine similarity of embedded article titles, ACM FAccT is, on average, most similar to HCOMP today (see \autoref{tab:nearest_conf}).
    Examples of works published at HCOMP include the work by \citeauthor{Lage_Chen_He_Narayanan_Kim_Gershman_Doshi-Velez_2019} on factors that make machine learning models interpretable by humans~\cite{Lage_Chen_He_Narayanan_Kim_Gershman_Doshi-Velez_2019},
    \citeauthor{Ray_Yao_Kumar_Divakaran_Burachas_2019}'s work
    on evaluating the efficacy of explanations in human-AI collaborative tasks \cite{Ray_Yao_Kumar_Divakaran_Burachas_2019},
    and 
    \citeauthor{Hase_Chen_Li_Rudin_2019}'s
    work on interpretability of vision models with hierarchical prototypes     \cite{Hase_Chen_Li_Rudin_2019}.
%
% <NEED TO EXPLAIN WHAT IS DIFFERENT ABOUT THE LATEST INTERPRETABILITY/EXPLAINABILITY STUDIES AT HCOMP>
% SOME CONCLUSION WHY THIS IS NOVEL AND POTENTIALLY BREAKING WITH PRIOR PARADIGM - e.g., is this even crowdsourcing research?
% Fairness and bias have, of course, been topics of study at HCOMP in the past.
These examples suggest that interpretability and explainability have emerged as novel themes at HCOMP, reflecting a broadening of the community's scope beyond traditional crowdsourcing paradigms.

% ---
% \begin{figure*}[!htb]%
% \centering
% \includegraphics[width=.8\textwidth]{figures/conference-similarity.pdf}%
% \caption{Pairwise comparison of the HCOMP Conference with six related conferences, in terms of cosine similarity between article title embedding centroids.}%
% \Description{figure description}%
% \label{fig:similarity}%
% \end{figure*}%
%
%
\begin{table*}[!htb]%
\caption{Mean similarity between HCOMP (250~articles) and related conferences, by cosine similarity of centroids of article title embeddings}
\label{tab:nearest_conf}
\small
\begin{tabular}{lccccccccccccc}
\toprule
Conference & N & 2013 & 2014 & 2015 & 2016 & 2017 & 2018 & 2019 & 2020 & 2021 & 2022 & 2023 & 2024 \\
\midrule
CSCW   & 3081 & \textbf{0.671} & 0.670 & 0.623 & 0.660 & 0.706 & \textbf{0.743} & 0.657 & 0.748 & 0.603 & 0.688 & 0.637 & 0.750 \\
FAccT  & 657 & -- & -- & -- & -- & -- & -- & 0.714 & 0.623 & 0.618 & 0.791 & 0.723 & \textbf{0.805} \\
CI     & 220 & --* & --* & \textbf{0.737} & --* & \textbf{0.807} & 0.736 & 0.654 & 0.611 & \textbf{0.699} & 0.733 & \textbf{0.972} & 0.760 \\
IUI    & 612 & 0.566 & 0.611 & 0.553 & \textbf{0.664} & 0.617 & 0.635 & \textbf{0.771} & \textbf{0.752} & 0.679 & \textbf{0.811} & 0.744 & 0.742 \\
UMAP   & 232 & -- & -- & -- & 0.631 & 0.538 & 0.529 & 0.668 & 0.690 & 0.577 & 0.768 & 0.729 & 0.790 \\
AAMAS  & 2203 & 0.620 & \textbf{0.716} & 0.595 & 0.574 & 0.601 & 0.603 & 0.611 & 0.546 & 0.605 & 0.585 & 0.585 & 0.514 \\
\bottomrule
\end{tabular}
\\
{%
\raggedright
{\footnotesize {*}~Website no longer available.}%
}%
\end{table*}%
% ---

% % ---
\begin{figure*}[!htb]%
\centering
\includegraphics[width=.8\textwidth]{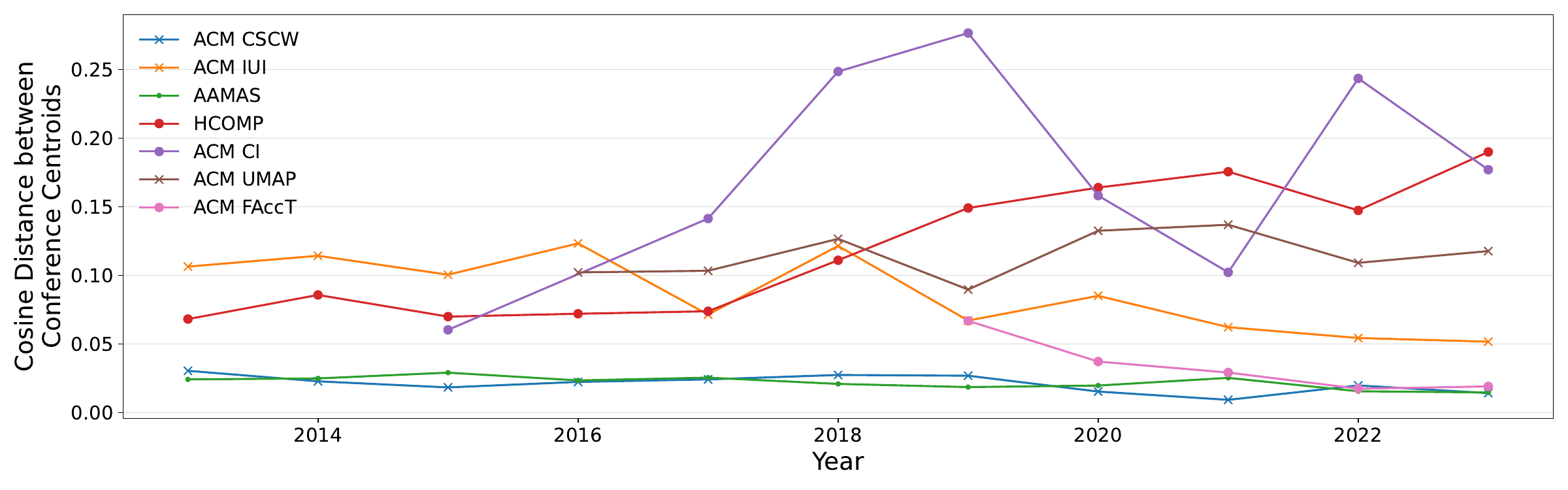}%
\caption{Cosine distances between centroids of article title embeddings in subsequent conference years at the HCOMP Conference (red) and six related conferences.
One can note that older venues, such as CSCW and AAMAS, display little centroid movement. This is also evident at ACM FAccT.
% , which since its creation, seems to have converged to a stable set of topics.
% The other venues, such as ACM CI and HCOMP, are more diverse in topics.
}%
\Description{todo}%
\label{fig:cosine-distances}%
\end{figure*}%

% \begin{figure*}[!htb]%
% \centering
% \includegraphics[width=.8\textwidth]{figures/cosine-distances-bars.pdf}%
% \caption{Cosine distances between subsequent conference years at the HCOMP Conference and six other conferences.}%
% \Description{todo}%
% \label{fig:cosine-distances}%
% \end{figure*}%

% \begin{figure*}[!htb]%
% \centering
% \includegraphics[width=.8\textwidth]{figures/conference-centroids.pdf}%
% \caption{\todo{Centroids}}%
% \Description{figure description}%
% \label{fig:similarity}%
% \end{figure*}%

% % ---

% ---
\begin{figure*}[!htb]%
\centering
\includegraphics[width=.85\textwidth]{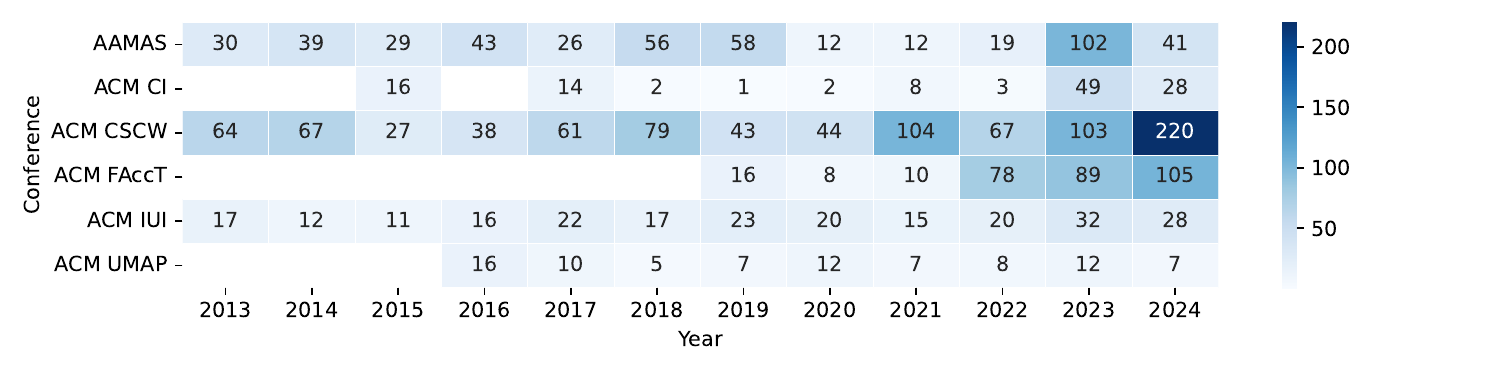}%
\caption{Shared keywords between the HCOMP conference and six related conferences over time (based on article titles)}%
\Description{figure description}%
\label{fig:shared-keywords}%
\end{figure*}%
% ---

\begin{figure*}[!htb]%
\centering
\includegraphics[trim=1.6cm 2cm 1.6cm 2cm,clip, width=.8\textwidth]{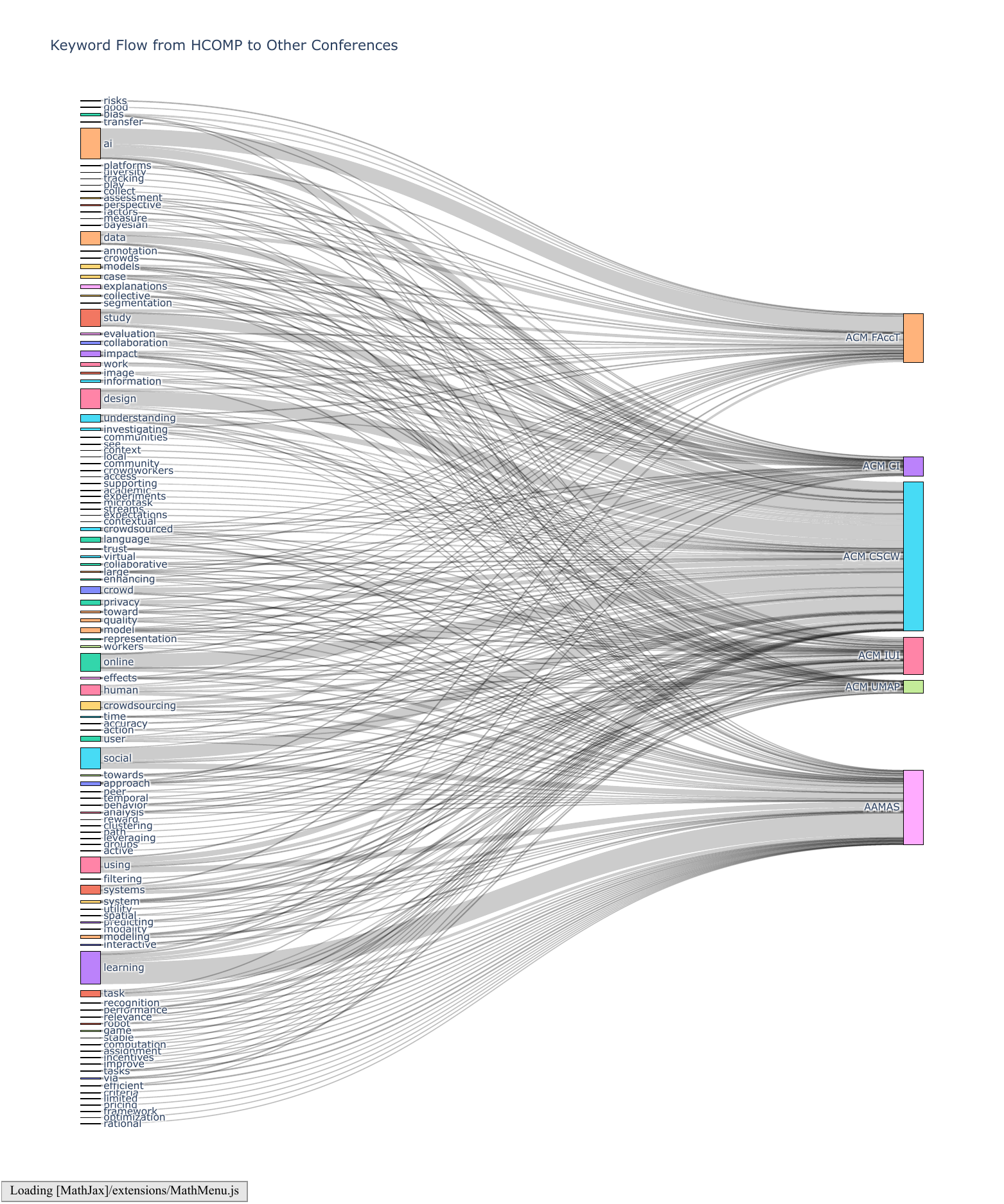}%
\caption{Shared keywords in the article titles at HCOMP (left) and six related conferences (ACM CI, ACM FAccT, ACM CSCW, ACM IUI, ACM UMAP, and AAMAS)}%
\Description{Sankey diagram showing that there are large numbers of shared keywords between HCOMP and six other conferences.}%
\label{fig:sankey}%
\end{figure*}

\begin{figure*}[htb]%
 \centering
 \begin{subfigure}[b]{.49\textwidth}%
        \centering
        \includegraphics[width=\textwidth]{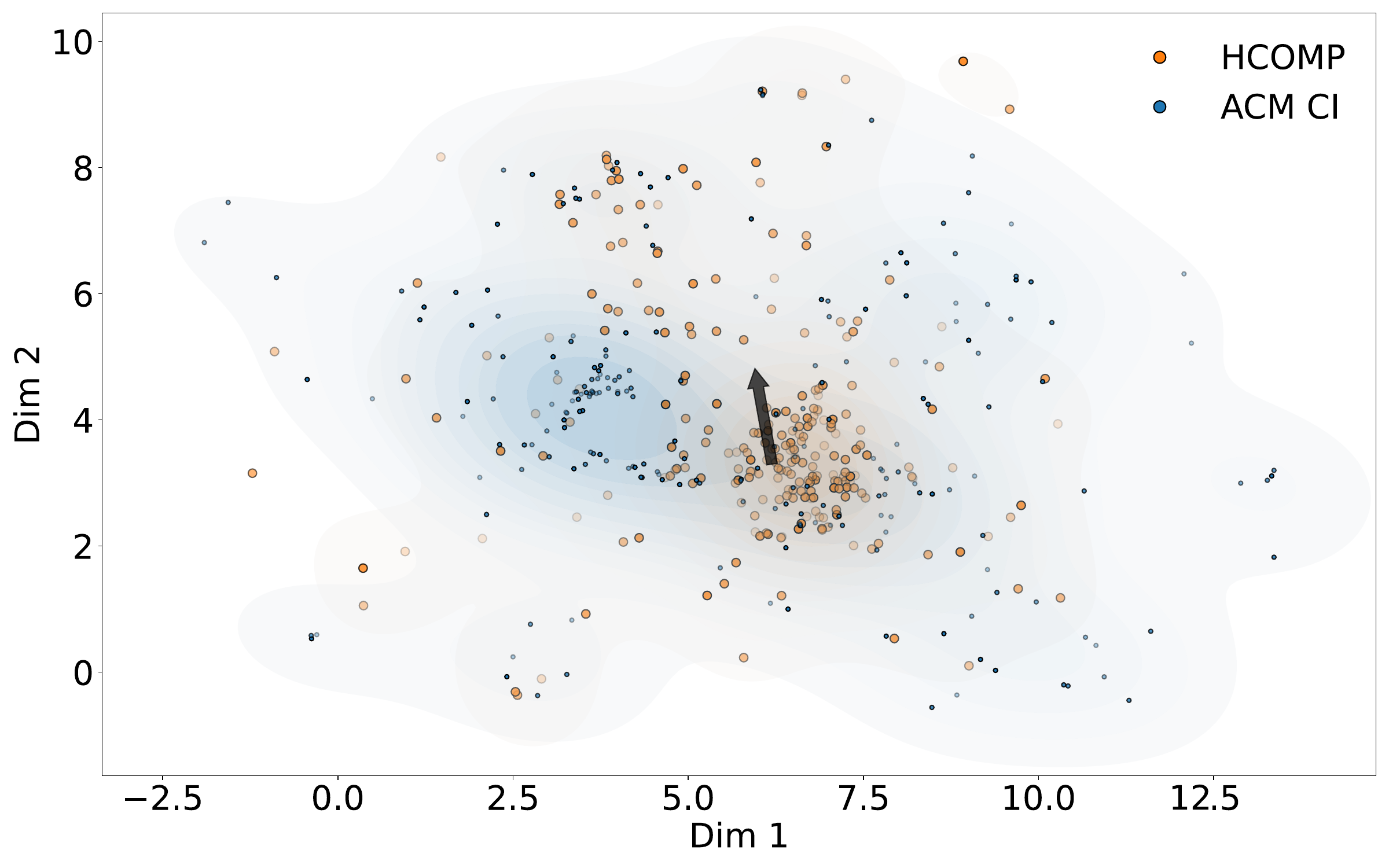}
        \caption{HCOMP and ACM CI}
    \Description{TODO}
        \label{fig:conf:a}
 \end{subfigure}
 \hfill
 \begin{subfigure}[b]{.49\textwidth}%
        \centering
        \includegraphics[width=\textwidth]{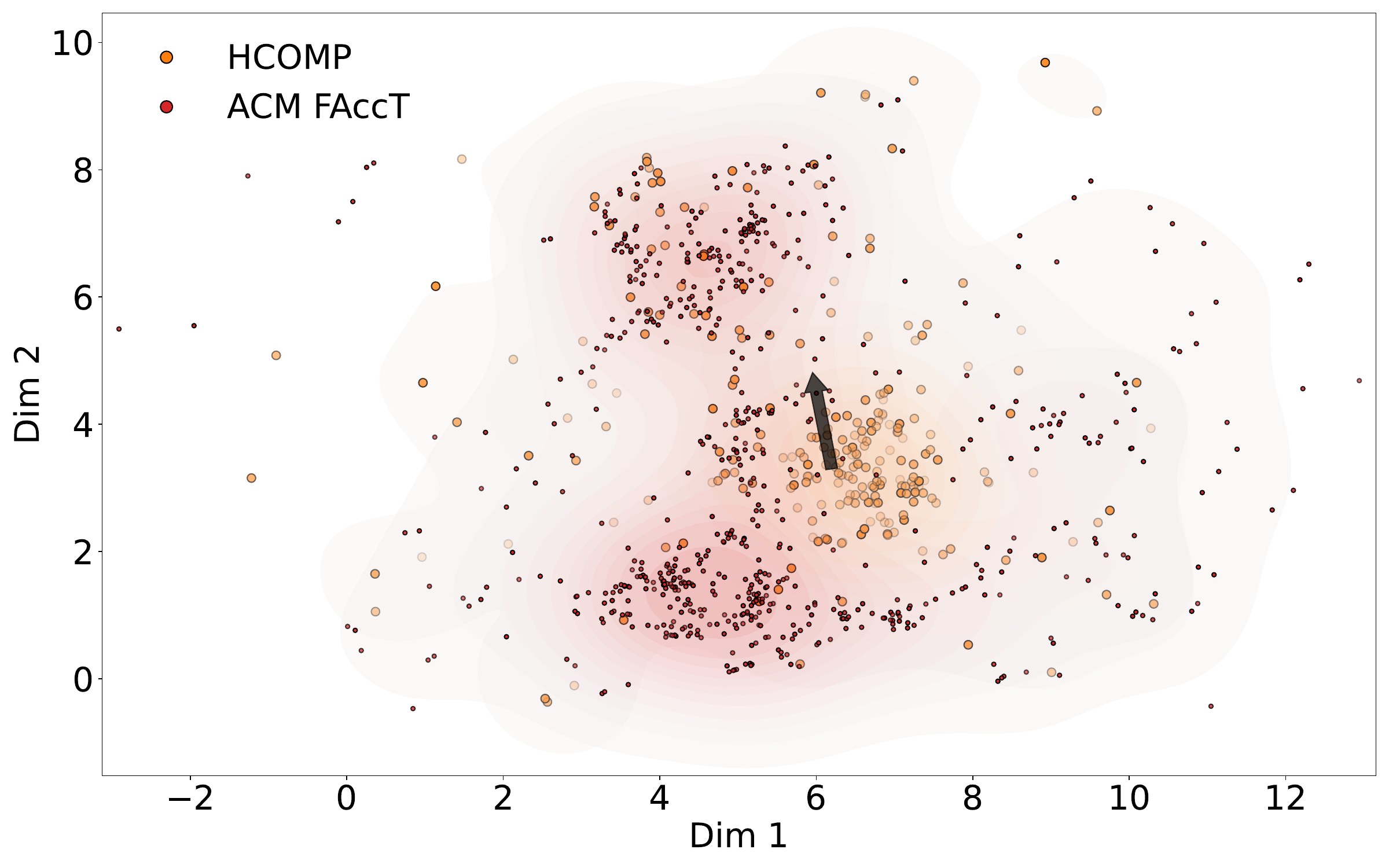}
        \caption{HCOMP and ACM FAccT}
    \Description{TODO}
        \label{fig:conf:b}
        \label{fig:conf:FAccT}
 \end{subfigure}
\\[.25\baselineskip]
 \begin{subfigure}[b]{.49\textwidth}%
        \centering
        \includegraphics[width=\textwidth]{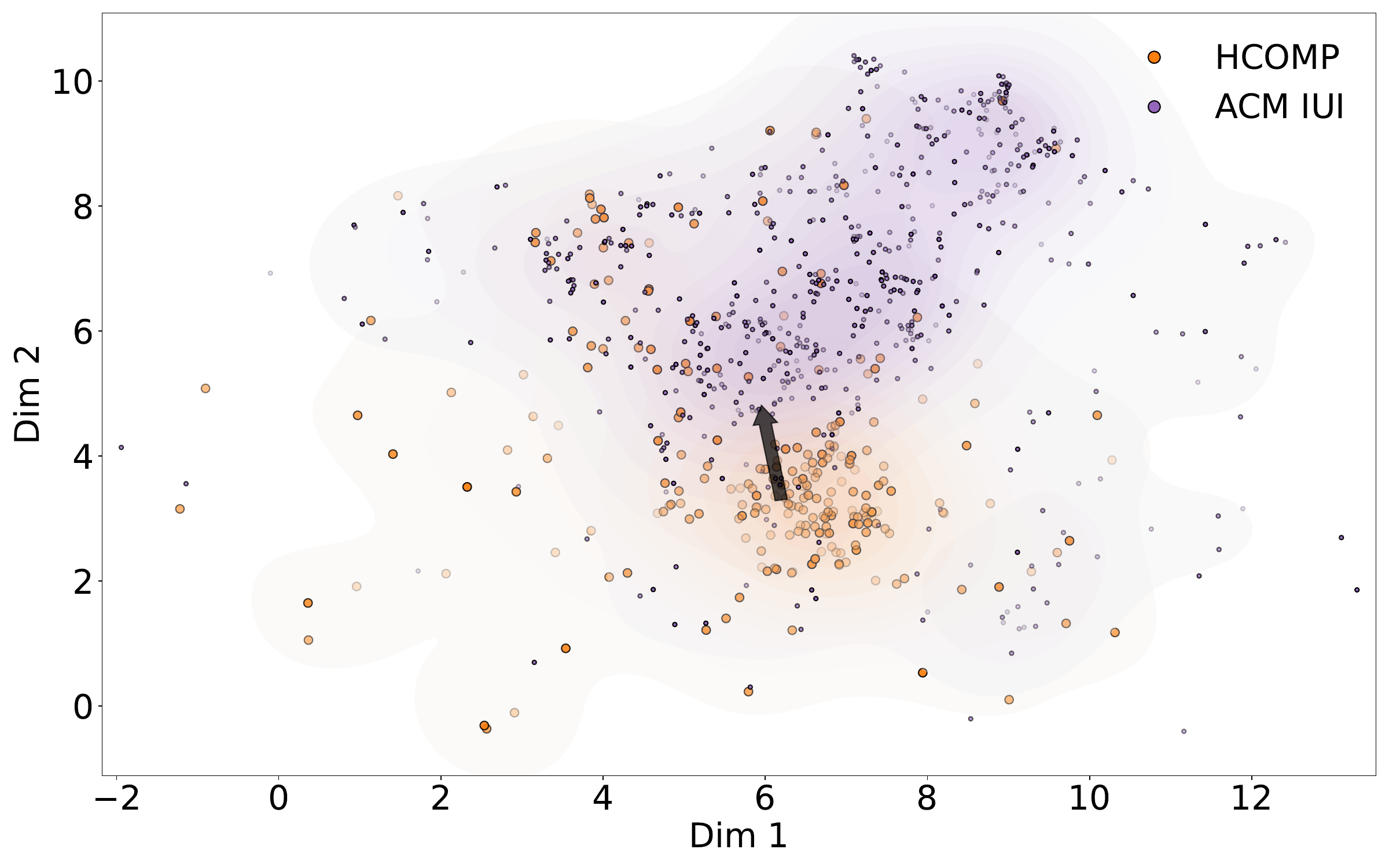}
        \caption{HCOMP and ACM IUI}
    \Description{TODO}
        \label{fig:conf:c}
 \end{subfigure}
 \hfill
 \begin{subfigure}[b]{.49\textwidth}%
        \centering
        \includegraphics[width=\textwidth]{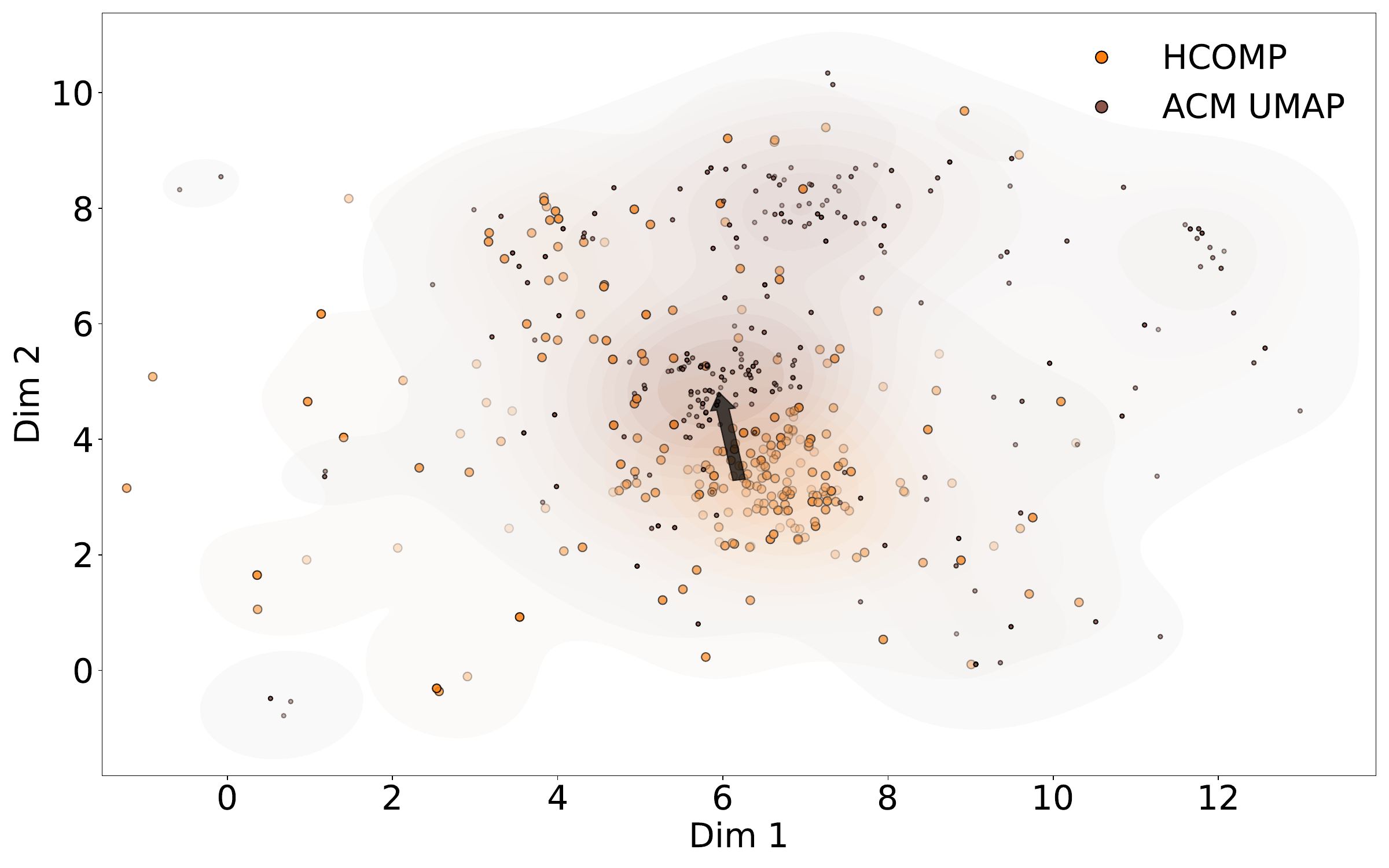}
        \caption{HCOMP and ACM UMAP}
    \Description{TODO}
        \label{fig:conf:d}
 \end{subfigure}
\\[.25\baselineskip]
 \begin{subfigure}[b]{.49\textwidth}%
        \centering
        \includegraphics[width=\textwidth]{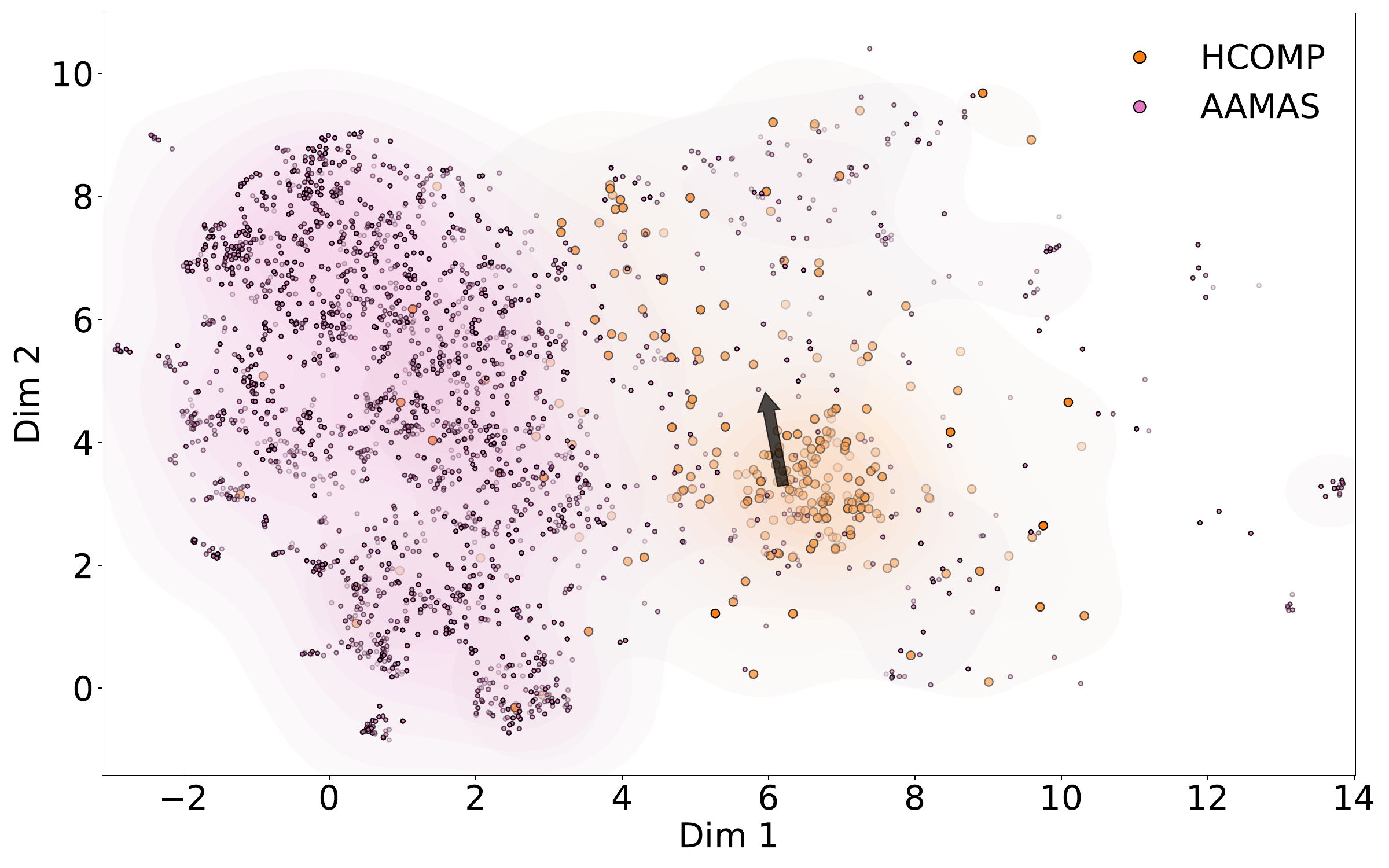}
        \caption{HCOMP and AAMAS}
    \Description{TODO}
        \label{fig:conf:e}
 \end{subfigure}
\hfill
 \begin{subfigure}[b]{.49\textwidth}%
        \centering
        \includegraphics[width=\textwidth]{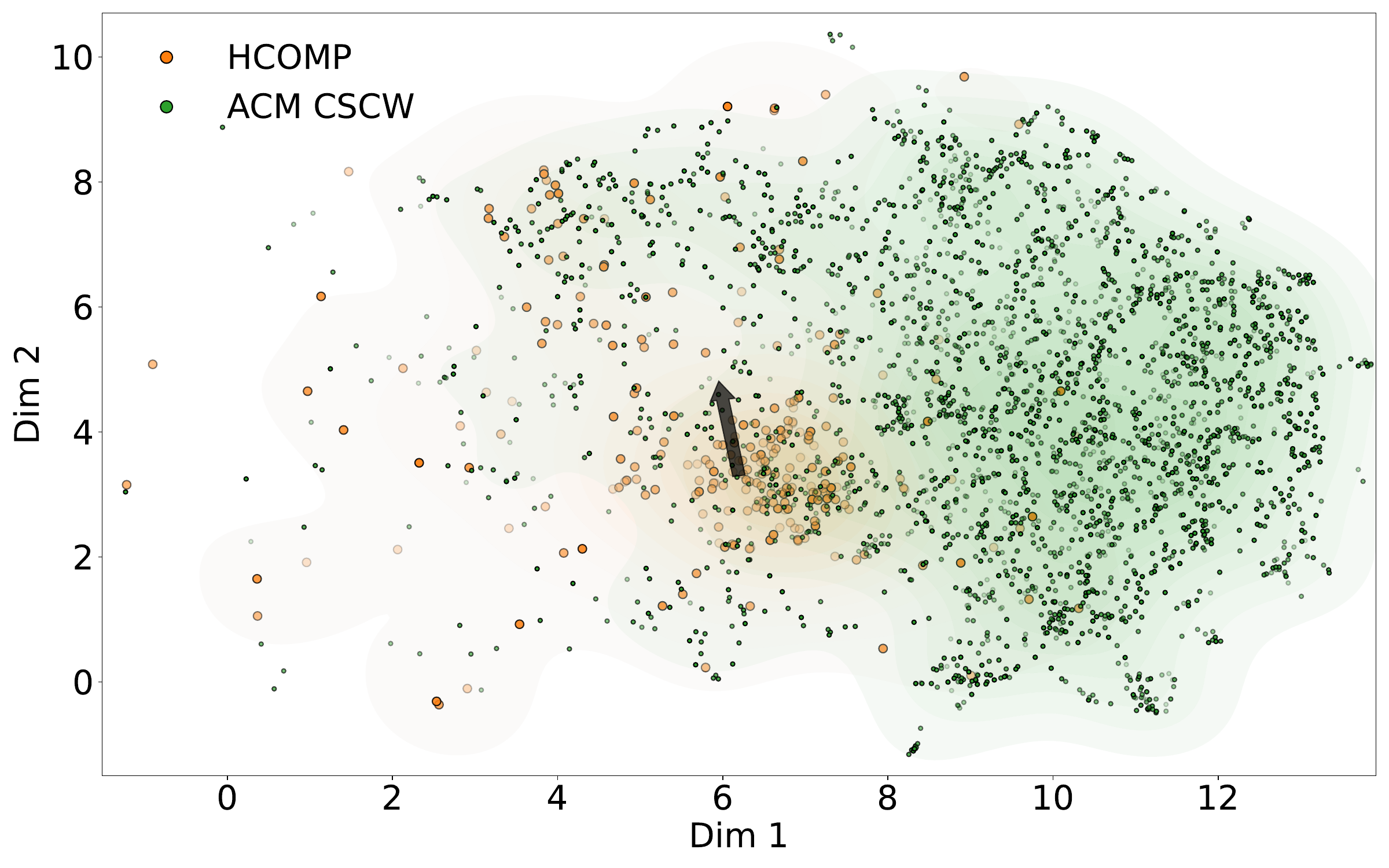}
        \caption{HCOMP and ACM CSCW}
    \Description{TODO}
        \label{fig:conf:f}
 \end{subfigure}
    \caption{The relation of HCOMP to six related conferences. Recent articles are more opaque. The direction of the HCOMP conference from 2013 to 2024, as depicted in \autoref{fig:teaser}, is annotated with an arrow.}%
    \Description{TODO}%
    \label{fig:conf}%
\end{figure*}

As the field of HCOMP evolved over the years, a growing similarity can also be noted with other conferences, in terms of centroid distances of article title embeddings (see \autoref{tab:nearest_conf}) and shared keywords (see \autoref{fig:shared-keywords} and \autoref{fig:sankey}).
% However, several other conferences are, today, also closely related, including ACM CI, ACM IUI, and ACM UMAP.
There is also a strong overlap in relevant keywords in article titles between HCOMP and 
% other conferences (see \autoref{fig:shared-keywords} and \autoref{fig:sankey}), with 
ACM CSCW (see \autoref{fig:shared-keywords}). % having the highest number of shared keywords in 2024.
Recently, HCOMP has also moved closer to the research spaces of ACM IUI, with its intelligent user interfaces providing a point of interaction between humans and AI, and ACM UMAP, which explores user modeling and personalization as a foundation for adaptive human-AI systems (see \autoref{fig:conf}).
% Plotting HCOMP's shift in relation to other conferences in \autoref{fig:conf}, we find empirical evidence that HCOMP has shifted, on average, to topics studied at ACM IUI and ACM UMAP, and
However, on average, HCOMP remains closely related to ACM CI (see \autoref{fig:conferences}).

% ---
% ---

\begin{figure*}[!htb]%
\centering%
  \includegraphics[width=.7\textwidth]{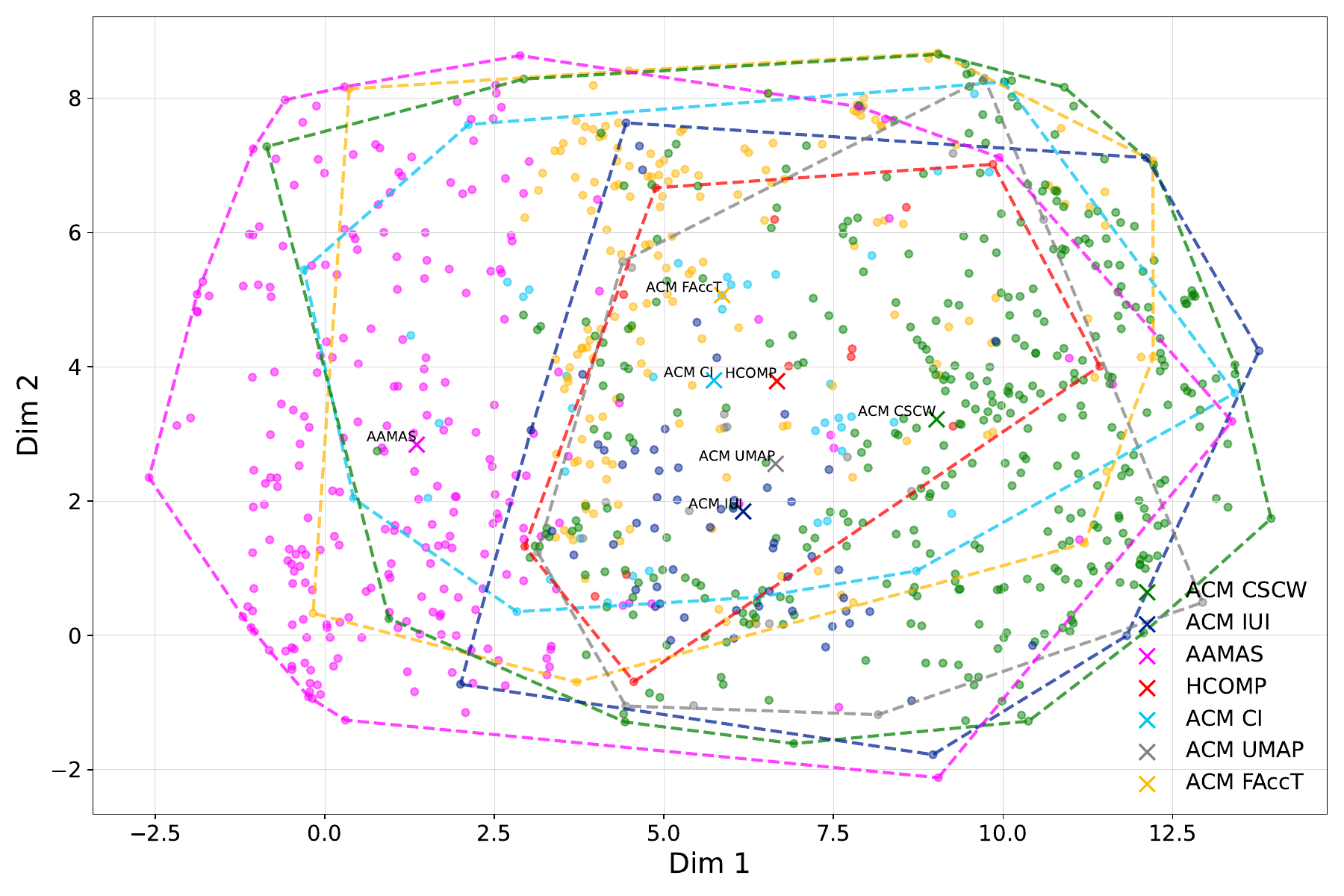}%
  \caption{The HCOMP conference's relation to six related conferences in 2024. Dots represent embeddings of article titles published in 2024. Convex hulls enclose all articles of a conference. The conference centroids are marked with X.}%
  \Description{HCOMP compared to six related conferences in 2024}%
  \label{fig:hulls}%
\end{figure*}

In all fairness, some overlap exists between all investigated conferences, as depicted in \autoref{fig:hulls}. Our analysis of centroids can only be an approximation of how conferences, as a whole, have developed over time.
It is interesting to note that some conferences---in particular CSCW, AAMAS, but also FAccT---demonstrate very little year-by-year centroid movements (see \autoref{fig:cosine-distances}), which may speak to the stability of research topics at these conferences.
% \todo{Low cosine distances could mean the conference venue has `stabilized' around a set of topics.}
% HCOMP and CI, on the other hand, have shifted topics more often and more diverse with different research streams being investigated in parallel.
%
%
 % there have been conferences with a strong overlap in research topics (see \autoref{tab:nearest_conf}) and keywords (\autoref{fig:shared-keywords} and \autoref{fig:sankey}), with HCOMP authors also publishing articles in several other related conferences.
    % For instance, the CSCW Conference had several research tracks related to crowdsourcing in the early years of HCOMP.
% Some works published at HCOMP relate to topics studied at the venues ACM IUI and ACM UMAP, for instance conversational systems \cite{x,x,x} and conversational agents \cite{x,x,x}.
%
% \autoref{fig:conf} depicts that HCOMP remains closely related to ACM Collective Intelligence.
% However, the recent migration by some authors moved the conference closer to topics studied at ACM FAccT (c.f. \autoref{fig:conferences}).
%
%
% \autoref{fig:conferences} depicts the recent migration of some authors at HCOMP toward topics that fit well with ACM FAccT.
{%
In the following section, we investigate whether a ``Gestalt-shift'' has taken place at the HCOMP conference.
}%

% ---
%
% \begin{figure}[!thb]
%   \includegraphics[width=\textwidth]{figures/conference-comparison-cscw.pdf}
%   \caption{%
%   Comparison of paper titles in AAAI HCOMP, ACM Collective Intelligence (CI) and ACM Conference on Fairness, Accountability, and Transparency (FAccT).
%   Older titles are more transparent.
%   }%
%   \Description{figure description}
%   \label{fig:conferences}
% \end{figure}
%
%
%
%
%
% \begin{figure*}[!thb]
% \centering
%   \includegraphics[width=\textwidth]{figures/conference-comparison-HCOMP,CI,FAccT,CSCW,IUI,UMAP,AAMAS.pdf}
%   \caption{%
%   \todo{%
%   Comparison of paper titles in AAAI HCOMP, ACM Collective Intelligence (CI), ACM Conference on Fairness, Accountability, and Transparency (FAccT), ACM CSCW, ACM IUI, and ACM UMAP.
%   }%
%   }%
%   \Description{figure description}
%   \label{fig:conferences:all}
% \end{figure*}
% ---

% ---
\subsection{HCOMP's Gestalt-Shift}
\label{sec:gestalt-shift}
\begin{figure*}[!htb]%
\centering%
  \includegraphics[width=.7\textwidth]{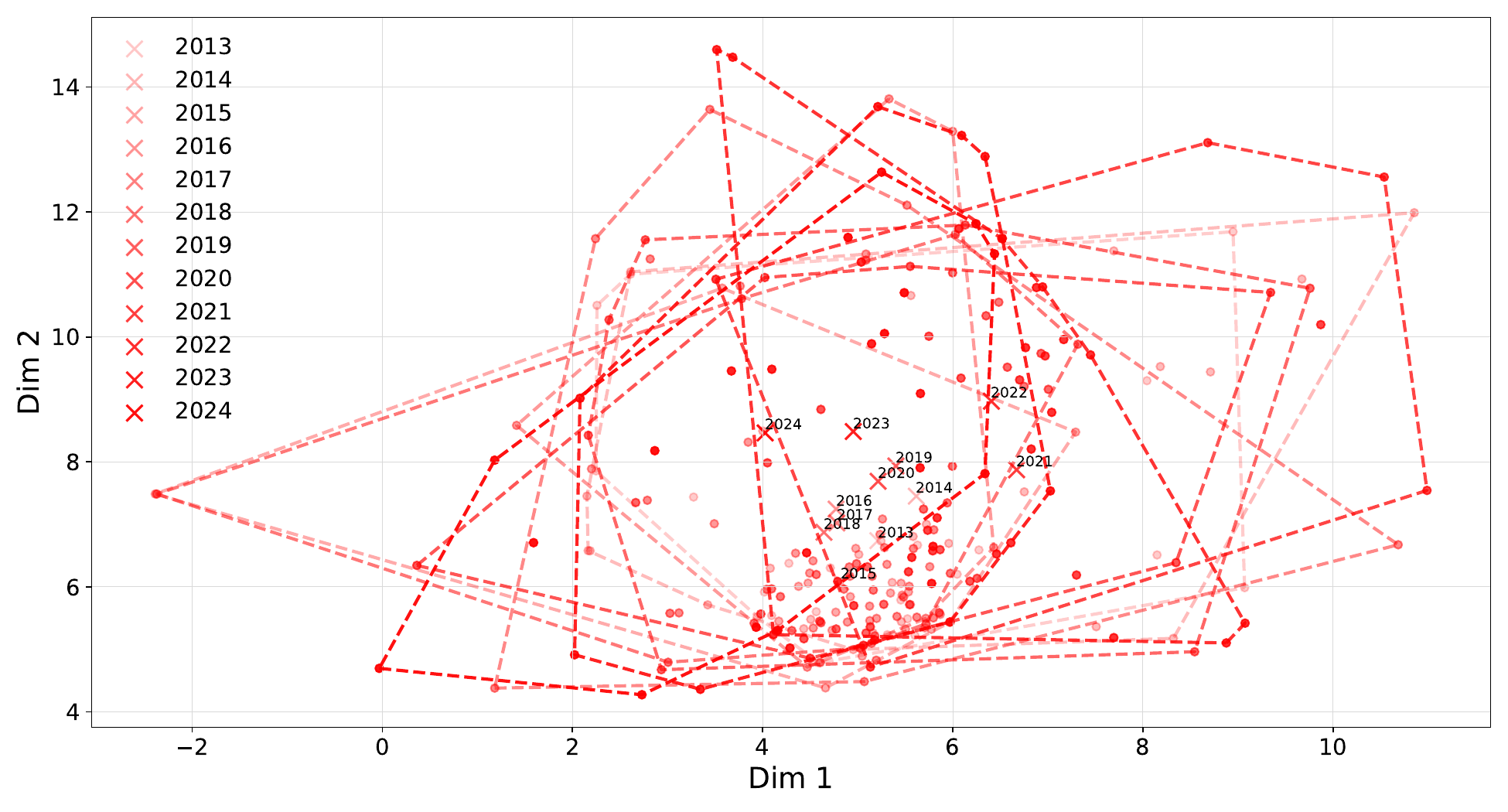}%
  \caption{Shifts at the HCOMP conference (2013--2024), as expressed through article title embeddings (represented as dots) and centroids for each year (marked with X). Convex hulls enclose articles in a given year.}%
  \Description{figure description}%
  \label{fig:gestalt}%
\end{figure*}

% ----------------------
Since around 2018, the HCOMP conference has been gradually moving away from its original research topics (see \autoref{fig:teaser} and \autoref{fig:conf}).
However, no sudden ``Gestalt-shift'' can be noticed in our analysis of centroid movements for the HCOMP conference (see \autoref{fig:cosine-distances} and \autoref{fig:gestalt}).
In these two figures, we would expect to see a large ``jump'' if a paradigm shift had taken place, yet the year-by-year movement of centroids is evolving gradually.
This may be evidence for the field still being in a phase of gradual transition, where some authors clearly switch to different topics, seeking alternatives to the
    persistent
    % irreconcilable
    % emerging
anomalies, while others continue to conduct ``normal science.''

% ---
% ---

Nevertheless, the fundamental assumptions of the field are increasingly being questioned.
This could indicate that the field has moved from ``normal science'' into the crisis phase of Kuhn's model, where anomalies and disturbances (e.g., the long-standing issues of quality in crowdsourced work, but also external shocks such as the introduction of large language models and technological advances in generative AI) accumulate, and the fundamental assumptions of the field are upended.
While the introduction of large language models has accelerated this for some authors, leading them to explore alternatives in topical areas that have traditionally not received much attention at HCOMP, the HCOMP conference seems to have, on average, not yet entered the revolutionary paradigm shift.%, as of yet.

% ======================
\section{Discussion}%
\label{sec:discussion}%
% ======================

% ----------------------
\subsection{Topical Shifts at HCOMP}
\label{sec:discussion:topics}
% ----------------------

%%% TOPIC SHIFT
The HCOMP conference originated from the Human Computation Workshop % , which was held four times 
before evolving into a stand-alone conference in 2013.
Since then, the HCOMP conference has gradually evolved and broadened its focus in recent years, to include critical perspectives at the intersection of humans and technology, % incorporating
touching on key topics from other conferences, such as ACM FAccT, IUI, and UMAP.
This shift in focus is reflected in the types of problems studied and % also
    in the terminology and conceptual framings that have become more prominent at the HCOMP conference.
While earlier years were focused on studies optimizing crowd workflows and task design, the trend since about 2018 highlights a renewed focus on integrating AI into socio-technical systems and the implications this has for human agency, fairness, and transparency.
Notably, research at HCOMP now engages with the design and evaluation of systems that involve humans and AI as collaborative agents, with a new emphasis on trust, explainability \& interpretability, and the responsible use of automation.
    For instance, the theme for the 2024 HCOMP conference was `\textit{Responsible Crowd Work for Better AI}.'
The disappearance of traditional co-word pairs, such as task--worker and crowd--worker, in favor of combinations such as human--AI and AI--crowd also reflects this conceptual broadening.
Rather than viewing the crowd as a passive labor pool, the field now increasingly investigates humans as active collaborators in systems shaped by algorithmic logic.

This reorientation suggests a departure from purely instrumental framings of human computation toward richer, more nuanced understandings of human–AI configurations.
In addition, the % alignment
overlap with neighboring conferences %--including FAccT, IUI, and UMAP--
    illustrates a changing scientific landscape % of interest 
    at HCOMP.
    % The coalescence around shared keywords and decreasing centroid distances between conferences
    % suggesting a convergence of interests across formerly distinct domains.
    % At the same time, the persistence of stable centers in other conferences, such as CSCW and AAMAS, underscores that this convergence is not universal but context-dependent.
%
%
Our findings suggest that HCOMP is undergoing a gradual redefinition of its intellectual boundaries.
Rather than abandoning its roots in crowd work and human computation, the community appears to be integrating these origins into a broader agenda that reflects contemporary concerns around AI ethics, collaboration, and human-centered design.
{In the following section, we discuss whether a paradigm shift has taken place at HCOMP.%
% and whether the field is in Kuhn's crisis phase.
}%
%
%
%
% ----------------------
\subsection{A Paradigm Shift at HCOMP?}%
\label{sec:discussion:results}%
% ----------------------
Has there been a revolutionary paradigm shift at HCOMP?
In recent years, HCOMP has observed an application and reinvention of methods and concepts to adapt to the new advances that generative AI has brought about. Examples include the application of workflow design---an area well-studied and refined in the crowdsourcing research community---to new human-AI configurations, and the increased focus on conversational agents, hybrid human-AI decision making, and human-AI teaming.
%This is reflected by the topical shift and re-focusing within the community in recent years.
If a paradigm shift had indeed already taken place, there would be some ``incommensurability'' between paradigms (i.e., they could not be directly compared since they use different methods or metrics for evaluation).
The adoption of the new paradigm would resemble a ``Gestalt-switch'' (i.e., a rather sudden perceptual switch in what the community identifies with rather than a gradual one).
Our investigation shows that shifts at HCOMP have been gradually occurring since about 2018, and there is likely no incommensurability between HCOMP's period of normal science and its current research.
We conclude that recent changes in HCOMP do not (at least yet) constitute a paradigm shift as per Kuhn's model.

The follow-up question, then, is whether HCOMP is in a phase of crisis.
The traditional HCOMP paradigm focused on effectively integrating and optimizing human computation, which provided ample opportunities for the field to make progress during a period of ``normal science.''
However, recent developments in AI can be argued to mean the field of human computation and crowdsourcing has entered the crisis stage in Kuhn's model, in which people have begun to fundamentally question and even undermine the role of ``human input'' in the age of generative AI.
Examples include the generation of data that would traditionally be collected from humans~\cite{hamalainendata,10.1145/3586183.3606763,wang-etal-2023-humanoid,park2024generativeagentsimulations1000} or commentaries on how language models can augment or replace human labor % in design tasks
\cite{schmidt2024simulating,10.1145/3712068}.
% These recent challenges---accelerated by the disruptive influence of generative AI---suggest that HCOMP may be experiencing a crisis phase, \todo{or entering it in the coming years}.
These recent challenges---accelerated by the disruptive influence of generative AI---% strongly
indicate that HCOMP is no longer operating within a stable period of ``normal science.''
Instead, the field appears to be entering the crisis phase of Kuhn’s model. 
Core assumptions that have long underpinned human computation---such as the unique value of human-generated data---are being actively questioned or undermined.
Researchers have begun to reorient their work toward issues of fairness, interpretability, and human-AI collaboration, and are increasingly publishing research that aligns more closely with the agendas of neighboring conferences like ACM FAccT, IUI, and UMAP.
While a full paradigm shift may not yet have occurred, the evidence suggests that HCOMP is now in the midst of a profound transformation in its epistemic foundations and research priorities.
%
%
% The impact of generative AI reaches beyond immediate concerns, and the community has of course embraced its new opportunities. 
% %While some concerns in the field have been always present, ``anomalies'', as per Kuhn's model, have now accumulated, accelerated by the technological advances in AI.
% %This phase could, eventually, lead to a revolutionary transformation in the field.
% In other words, some researchers in the research community are shifting their focus to other areas.
% %, finding new or alternative research communities as a result.
% %thereby essentially abandoning the research area of HCOMP.
% We found some support for this in the topics studied by some authors at HCOMP, which would fit well with ACM FAccT, ACM IUI, and ACM UMAP.
%
%
Related to this is the question of whether the community is shedding its old identity in this process.
% In the following section, we discuss the future of HCOMP in relation to these conferences.
In the following section, we speculate on the future of HCOMP in relation to other fields.

% ----------------------
\subsection{The Future of HCOMP}
\label{sec:discussion:future}
% ----------------------
% While we cannot make predictions on what will happen to the research area of HCOMP, 
We believe that by looking at the past, we can understand the present and critically inform decision-making for the future.
We have empirically identified that research at the HCOMP conference has gradually shifted focus since around 2018,  with an innovative reorientation from `workers' to `humans' taking place since 2021.
The research area has evolved, with notable shifts in research topics toward the intersection of AI and humans.
The field of HCOMP seems to have moved on from some of its past motor themes and is in a process of reorienting itself in terms of topics studied.

% Ignoring its many flaws and the skewed power dynamics between requesters and workers,
    Paid
    % Microtask
crowdsourcing was enormously important and instrumental to the revolution of artificial intelligence that we bask in today.
    For instance, early computer vision models relied on crowd workers labeling images and instruction fine-tuning via reinforcement learning from human feedback (RLHF) contributed to the flourishing of large language models.
Increasingly, however, companies turn to collect data with alternative, more cost-effective means---often for free---by spinning their own ``data flywheels'' without the need for outsourcing human labor. % drawing on free human labor.
    Tesla, for instance, collects massive amounts of data from Tesla vehicles in-the-wild. 
    OpenAI uses conversations and user feedback for training future generations of their chatbots.
    And social media companies---at least until recently---maintained large numbers of in-house content moderators \cite{BBC,mess}.
However, with the emergence of now ubiquitous free large language models and technological advances in automation, the human contribution in crowdsourcing is called into question.
%%% Social media companies, like Facebook and TikTok, employed large amounts of in-house  moderators (although they have moved away from this recently).
% On the other hand, moderation on social platforms remains a critical issues, with companies employing hords of internal moderators.
% Although, both X (formerly Twitter) and Meta's platforms have announced moves toward less moderation.

One question during this phase of transition is whether HCOMP should remain its own research field, or merge with another conference.
Given the strong interest of researchers in studying large language models---some even advocating for replacing human participants \cite{schmidt2024simulating,park2024generativeagentsimulations1000,wang-etal-2023-humanoid,10.1145/3586183.3606763}---there is an increasing overlap of topics studied at HCOMP and other conferences.
% We identified several conferences that are---as a whole---
We found empirical evidence that some researchers have moved closer to topics studied at ACM FAccT (\autoref{fig:conferences}), and HCOMP---as a whole---has moved closer to conferences such as ACM IUI and UMAP (\autoref{fig:conf}).
However, one of the closest conferences, in terms of topic similarity, remains the ACM Collective Intelligence Conference (see \autoref{fig:conferences} and \autoref{fig:conf}).
% However, the ACM Collective Intelligence Conference remains one of the most closely related conferences (see \autoref{fig:conferences}).
With its broadening topical focus, the HCOMP conference fits well together with CI.
This is, to some extent, no surprise, given that the two conferences were co-located and held jointly in the past.
%%% conclusion/ENDING
We argue there is still a place at HCOMP for research on crowdsourcing.
However, as evident in our work, some introspection and reflection on the past is needed to inform HCOMP's future,
and in the age of generative AI, % has implications for shifting the 
the purpose of human labor may need to shift from 
    % a generation-heavy purpose to a verification or oversight-centered purpose.
    data generation to verification and oversight.
Our work contributes data-driven insights to this discussion.%
% of the fundamental assumptions in the field should be thought over and, perhaps, 
% abandoned to resolve some of the field's longstanding issues and 
% adapted to the new reality of ubiquitous AI.
%
%
%
%
% ----------------------
\subsection{Limitations and Future Work}%
\label{sec:discussion:limitations}%
% ----------------------
We acknowledge a number of limitations to our work.
First, with a mean of 20.8~articles published each year, HCOMP is a small conference, and there are only a limited number of data points each year.
This affects our analysis, in particular \autoref{fig:gestalt}.
Second, we acknowledge that the field of human computation and crowdsourcing research is much larger than just the HCOMP conference. However, we use the HCOMP conference
    % is our subject of study in this article
    as a proxy for the wider research field on human computation and crowdsourcing.
%We believe the observations hold true for the wider field of crowdsourcing research.
Future work could extend the analysis % to more venues 
to develop a deeper understanding of the impact of recent technological developments on the field of HCOMP.
%
% These terms are generic domain terms and appear frequently in many documents (high TF), but may not be very informative since they are common across the corpus (low IDF).
%
A related limitation is our use of embeddings, which encode % the
    semantic meaning of % the
    text.
There are limitations to interpreting these plots (see \cite{Understanding-UMAP,oppenlaender2023mapping}).
The complexity of human decision-making in research cannot fully be captured by embedding titles of published articles.
Further, in the interpretation of our findings, one needs to consider that HCOMP is a highly specialized venue, rich in % domain-specific
generic domain terms, such as `crowdsourcing' and `crowd work'.
This may have influenced the results.
% Our work, however, is a offers an approximation of trends in HCOMP, backed by an extensive review of the literature.
Future work could involve HCOMP researchers in a qualitative investigation to address these limitations.
Last, a limitation to our approach % to measuring the paradigm shift at the HCOMP conference 
    is that recent advances in AI can also be used to study existing research topics, by simply replacing existing methods. The Conference on Human Factors in Computing Systems (CHI), for instance, has seen a strong uptake in both studying and using large language models \cite{2501.12557.pdf}.
% Some authors have called this the ``lllm-ification'' of CHI \cite{2501.12557.pdf}. 
Such shifts are much harder to measure because in this case, research topics stay the same, and only methods change.
% Our approach would not be able to capture such shifts.
% Further, it is not clear if this constutes a paradigm shift, since the new `method' and `tool' of AI is simply being used to study existing topics.
Also, a natural drift in topics can be expected, moving fields away from their original topics. This would, however, not constitute a sudden incommensurable
paradigm shift.%
%
%
%
% ======================
\section{Conclusion}%
\label{sec:conclusion}%
% ======================
The Conference on Human Computation and Crowdsourcing is at a crossroads.
% , and its future is uncertain.
We found that research at the HCOMP conference has gradually shifted away from its traditional motor themes % research topics
toward artificial intelligence, explainability \& interpretability, conversational systems, and human-AI decision-making.
This could mean that HCOMP has transitioned---prompted by anomalies brought about by generative AI, challenging and undermining fundamental assumptions in the field---from  a period of ``normal science'' into a new phase.
However, we argue this shift cannot be called a revolutionary paradigm shift, according to Kuhn's framework, as of yet.
Instead, the field's research focus has gradually broadened to include critical perspectives at the intersection of humans and technology, incorporating topics from other conferences, such as ACM FAccT, ACM IUI, and ACM UMAP.
Ultimately, the fate of any given venue hinges on many factors outside the evolution of its topics, for instance funding and community spirit.
    % for instance, funding and community spirit, and recognition of conferences with respect to national or institutional regulations. 
% With our work, we % hope to
%     contribute a meaningful and data-informed % analysis
%     piece to this broader discussion. %
With our work % we hope to
    contribute a meaningful and data-informed piece to this broader discussion.%

\balance

%%%%%%%%%%%%%%%%%%%%%%%%%%%%%%%%%%%%
%%%%%%%%%%%%%%%%%%%%%%%%%%%%%%%%%%%%

% \begin{table}
%   \caption{Frequency of Special Characters}
%   \label{tab:freq}
%   \begin{tabular}{ccl}
%     \toprule
%     Non-English or Math&Frequency&Comments\\
%     \midrule
%     \O & 1 in 1,000& For Swedish names\\
%     $\pi$ & 1 in 5& Common in math\\
%     \$ & 4 in 5 & Used in business\\
%     $\Psi^2_1$ & 1 in 40,000& Unexplained usage\\
%   \bottomrule
% \end{tabular}
% \end{table}

% \begin{table*}
%   \caption{Some Typical Commands}
%   \label{tab:commands}
%   \begin{tabular}{ccl}
%     \toprule
%     Command &A Number & Comments\\
%     \midrule
%     \texttt{{\char'134}author} & 100& Author \\
%     \texttt{{\char'134}table}& 300 & For tables\\
%     \texttt{{\char'134}table*}& 400& For wider tables\\
%     \bottomrule
%   \end{tabular}
% \end{table*}

%   \caption{figure caption}
%   \Description{figure description}

% \begin{acks}
% \end{acks}

%TC:ignore
\bibliographystyle{ACM-Reference-Format}
\bibliography{main}
%TC:endignore

\end{document}